\documentclass[english]{IEEEtran}
\usepackage[T1]{fontenc}
\usepackage[latin9]{inputenc}
\usepackage{pifont}
\usepackage{float}
\usepackage{booktabs}
\usepackage{bbding}
\usepackage{ifsym}
\usepackage{amsmath}
\usepackage{amssymb}
\usepackage{graphicx}
\usepackage{setspace}
\usepackage{wasysym}
\PassOptionsToPackage{normalem}{ulem}
\usepackage{ulem}

\makeatletter

\providecommand{\tabularnewline}{\\}
\floatstyle{ruled}
\newfloat{algorithm}{tbp}{loa}
\providecommand{\algorithmname}{Algorithm}
\floatname{algorithm}{\protect\algorithmname}

\allowdisplaybreaks
\usepackage{bibunits}
\usepackage{amsthm}
\usepackage{amssymb}
\newtheoremstyle{boldstyle1}    
  {\topsep}                    
  {\topsep}                    
  {\itshape}                   
  {}                           
  {\bfseries}                  
  {.}                          
  {.5em}                       
  {}                           

\newtheoremstyle{boldstyle2}    
  {\topsep}                    
  {\topsep}                    
  {}                           
  {}                           
  {\bfseries}                  
  {.}                          
  {.5em}                       
  {}                           

\newtheoremstyle{proofstyle}    
  {\topsep}                    
  {\topsep}                    
  {}                           
  {}                           
  {\itshape}                  
  {.}                          
  {.5em}                       
  {}                           

\theoremstyle{boldstyle2}       
\newtheorem{definition}{Definition} 
\newtheorem{example}{Example}

\theoremstyle{boldstyle1}       
\newtheorem{proposition}{Proposition} 
\newtheorem{lemma}{Lemma}

\renewcommand{\qed}{\hfill\scalebox{0.7}{$\blacksquare$}}
\theoremstyle{proofstyle}       
\newtheorem*{proof_temp}{Proof}
\newenvironment{proof_dat}{\begin{proof_temp}}{\qed\end{proof_temp}}

\theoremstyle{proofstyle}       
\newtheorem{remark}{Remark}

\makeatother

\usepackage{babel}
\begin{document}
\title{The Mean of Multi-Object Trajectories }
\author{Tran Thien Dat Nguyen, Ba Tuong Vo, Ba-Ngu Vo, Hoa Van Nguyen, and
Changbeom Shim\thanks{The authors are with the School of Electrical Engineering, Computing
and Mathematical Sciences, Curtin University, Australia (email: \{t.nguyen1,
ba-tuong.vo, ba-ngu.vo, hoa.v.nguyen, changbeom.shim\}@curtin.edu.au).
Corresponding author: Changbeom Shim.}\thanks{This work was supported by the ARC Future Fellowship FT210100506 and
the Office of Naval Research Global N629092512046.}}
\maketitle
\begin{abstract}
This paper introduces the concept of a mean for trajectories and multi-object
trajectories (defined as sets or multi-sets of trajectories) along
with algorithms for computing them. Specifically, we use the Fr\'{e}chet
mean, and metrics based on the optimal sub-pattern assignment (OSPA)
construct, to extend the notion of average from vectors to trajectories
and multi-object trajectories. Further, we develop efficient algorithms
to compute these means using greedy search and Gibbs sampling. Using
distributed multi-object tracking as an application, we demonstrate
that the Fr\'{e}chet mean approach to multi-object trajectory consensus
significantly outperforms state-of-the-art distributed multi-object
tracking methods. 
\end{abstract}


\section{Introduction\protect\label{sec:Introduction}}

Multi-object estimation involves determining the trajectories of
an unknown, time-varying number of objects from noisy sensor measurements.
Formally, a\textit{ trajectory} is a function that maps time indices
from a discrete time window to elements in some state space, and a\textit{
multi-object trajectory} is a point pattern or a multi-set\footnote{A multi-set may contain repeated elements.}
of trajectories \cite{beard2020solution}. This general definition
covers trajectories that are fragmented (i.e., objects can disappear
and then reappear later), which is usually the case with the output
of most tracking systems. Estimating the multi-object trajectory lies
at the heart of many applications ranging from space exploration \cite{persico2018cubesat},
surveillance \cite{liu2022multi,nguyen2019trackerbots}, computer
vision \cite{ma2024track,zhang2021fairmot}, to cell biology \cite{mavska2023cell,nguyen2021tracking}. 

A fundamental concept in estimation (and statistics) is the mean or
average. Intuitively, the mean can be interpreted as the typical point
and is, arguably, the most widely used summary statistic. In the presence
of uncertainty, the mean is also useful as an estimator of the variable
of interest. Additionally, averaging is the most popular form of consensus--reaching
an agreement about an entity of interest--crucial for the fusion
of data/information from multiple sources. Consensus has a wide range
of applications in managerial science, statistics, engineering and
computer science \cite{winkler1968consensus,floyd1992managing,olfati2007consensus,xiao2005scheme}.
In computer science and engineering, consensus-based algorithms are
extensively used in distributed computing \cite{lynch1996distributed},
energy systems \cite{schiffer2015voltage}, communication networks
\cite{li2009consensus}, remote sensing \cite{benediktsson2003multisource},
sensor fusion \cite{xiao2005scheme}, networked and distributed estimation
\cite{luo2006distributed}, \cite{iyer2022wind}.

\begin{figure}
\begin{centering}
\includegraphics[width=0.98\columnwidth]{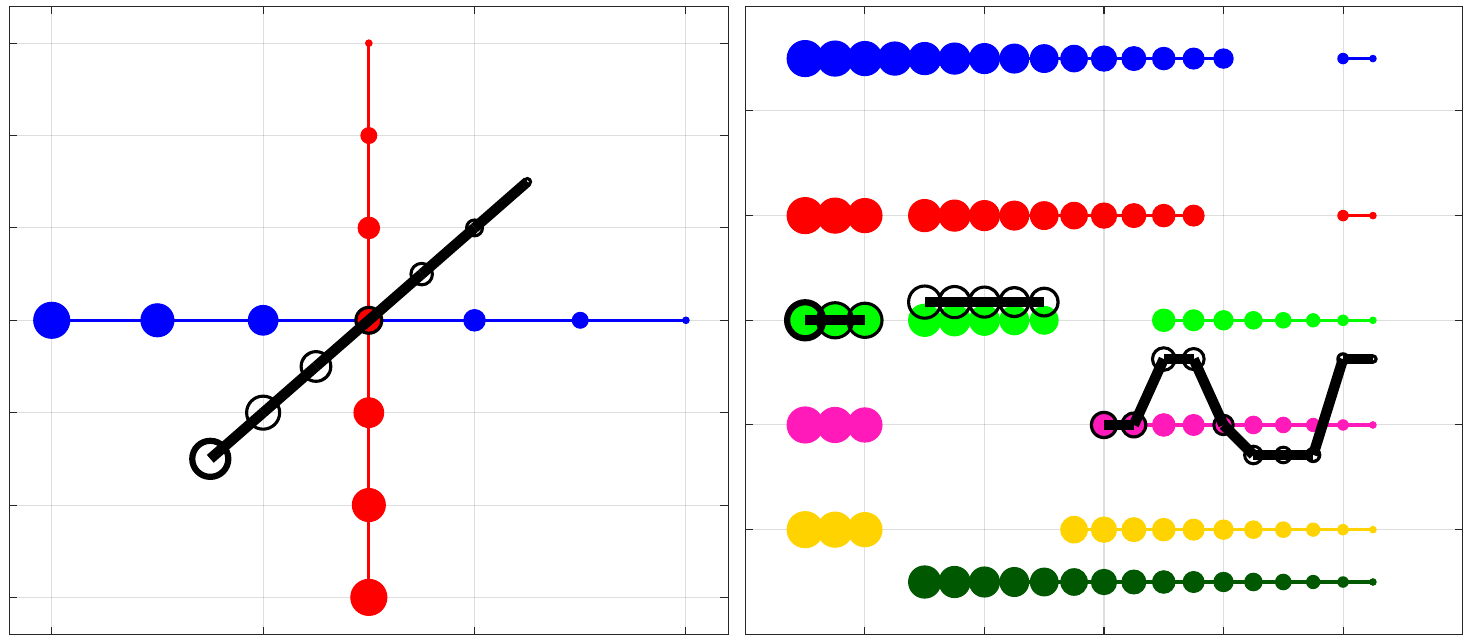}
\par\end{centering}
\caption{Two examples of the mean trajectory (thick black) of multiple input
trajectories (color). Marker sizes indicate times, with the same size
representing the same time, and sizes decrease toward the ends of
trajectories. Consecutive instances of a trajectory are connected
with solid lines. \protect\label{fig:samples-mean-so}\vspace{-0em}}
\end{figure}

\begin{figure*}
\begin{centering}
\includegraphics[width=0.9\textwidth]{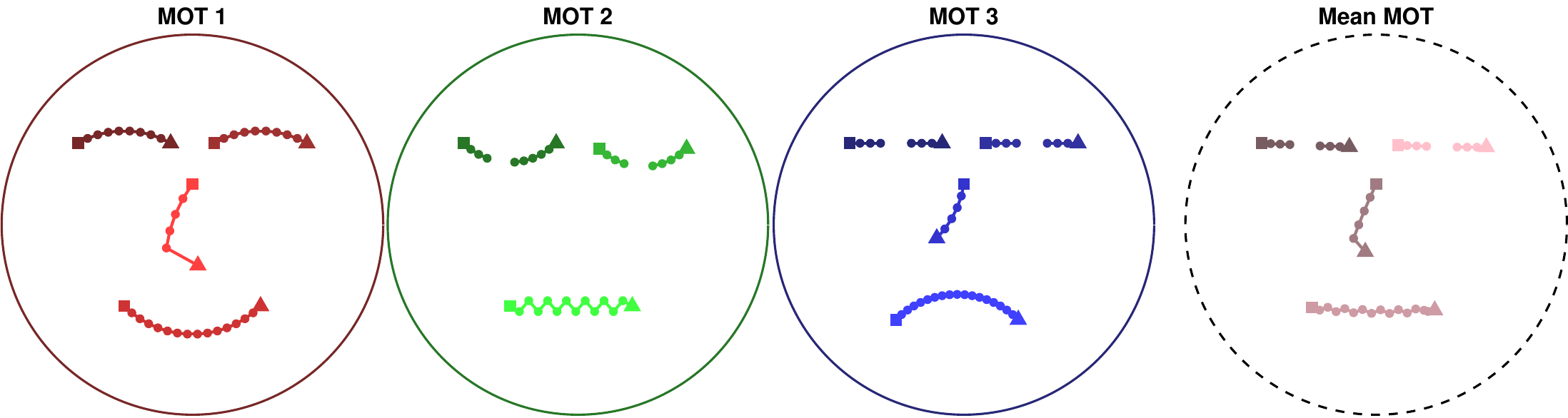}
\par\end{centering}
\begin{raggedleft}
\vspace{0.1em}\includegraphics[bb=0bp 20bp 638bp 74bp,clip,width=0.25\textwidth]{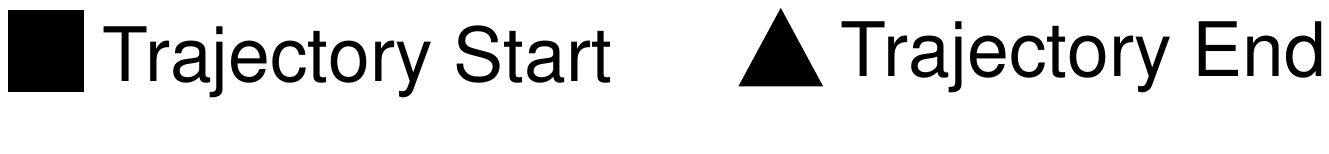}\thickspace{}\thickspace{}\thickspace{}\thickspace{}\thickspace{}\vspace{-0.95em}
\par\end{raggedleft}
\begin{centering}
\rule[0.5ex]{0.95\textwidth}{0.5pt}
\par\end{centering}
\begin{centering}
\vspace{-0.1em}\includegraphics[clip,width=0.9\textwidth]{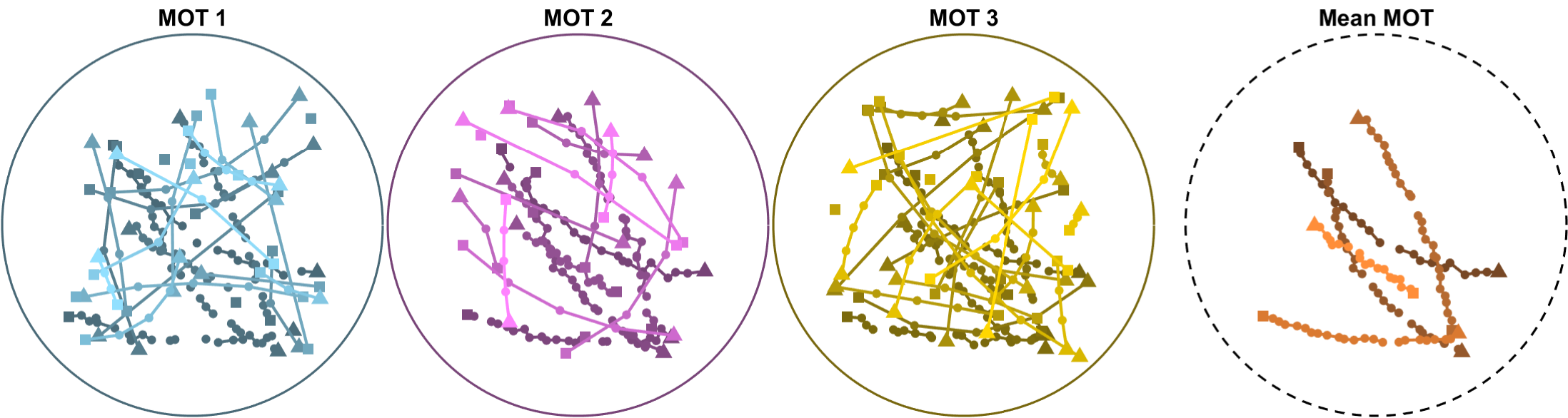}
\par\end{centering}
\caption{Examples of the mean multi-object trajectory (MOT) of three sample
multi-object trajectories. Each trajectory in a multi-object trajectory
has a distinct color. Consecutive instances of a trajectory are connected
with solid lines. Intuitively, the mean tends to smooth out the data
points. \protect\label{fig:samples-mean-mo}\vspace{-0.5em}}
\end{figure*}

While it is customary that the mean is simply the sum of the data
points divided by the number of data points, this is not the case
in multi-object estimation. When the data points are trajectories
of different lengths (see e.g., Figure \ref{fig:samples-mean-so}),
or multi-object trajectories (see e.g., Figure \ref{fig:samples-mean-mo}),
the customary notion of average is not applicable because addition
is not defined. The fundamental question then arises: what constitutes
an ``average'' trajectory or an ``average'' multi-object trajectory?
This question is crucial in multi-object estimation, yet, to the best
of our knowledge, there is no existing work addressing the concept
of an ``average'' trajectory or an ``average'' multi-object trajectory.

This work introduces the notion of average for trajectories and multi-object
trajectories via the Fr\'{e}chet mean, accompanied by algorithms for computing
them. The Fr\'{e}chet mean generalizes the concept of \textquotedblleft average\textquotedblright ,
\textquotedblleft centroid\textquotedblright{} or ``barycenter''
in Euclidean space to an arbitrary metric space \cite{lou2020differentiating}.
In the same way as the Euclidean mean captures the geometric relationship
induced by the Euclidean distance, the Fr\'{e}chet mean captures the geometric
relationship induced by the metric, allowing the physical interpretation
as the \textquotedblleft average\textquotedblright{} point. In a multi-object
setting, the \textit{optimal sub-pattern assignment} (OSPA) metric
\cite{schuhmacher2008consistent} is a meaningful distance between
trajectories as well as between multi-sets/point patterns of trajectories
\cite{beard2020solution}. OSPA-based distances and the concept of
Fr\'{e}chet mean enable the formulations of physically meaningful averages
for trajectories and multi-object trajectories. In summary, the key
contributions of this paper are:
\begin{itemize}
\item A formulation of the mean trajectory, and algorithms for computing
this mean with respect to (w.r.t.) the OSPA trajectory metric;
\item A formulation of the mean multi-object trajectory, and algorithms
for computing this mean w.r.t. OSPA-based multi-object trajectory
metrics.
\end{itemize}
Additionally, we demonstrate the capability of the proposed means
in achieving consensus, using networked multi-object tracking as a
specific area of application. Due to the proliferation of low-cost
sensors that enable the deployment of large networks capable of monitoring
large spatial regions, networked multi-object tracking is an emerging
trend \cite{iyer2022wind}. This approach leverages advances in communication
and sensing technologies to improve accuracy by fusing information
from multiple nodes in a flexible and reliable manner, resilient to
node failures \cite{luo2006distributed}. 

Fusion is used to determine a multi-object trajectory description
(e.g., estimates or probability distributions)�that best agrees with
the local multi-object trajectory descriptions at each node of the
network. The fused description can be posed as a consensus on the
local multi-object trajectories. The proposed Fr\'{e}chet mean is a logical
basis for consensus since by construction the mean is the closest
(in aggregate distance) to all inputs from individual nodes. Further,
results on the statistical consistency of Fr\'{e}chet means have been
established in \cite{schotz2022strong}. Indeed, as illustrated in
Figure \ref{fig:samples-mean-mo}, the Fr\'{e}chet mean effectively suppresses
noise and outliers. Notably, the proposed consensus can fuse node
outputs regardless of the types of local trackers, and shows significant
performance improvements over current state-of-the-art techniques.

The paper is organized as follows. Section \ref{sec:Background} provides
the background on averaging in multi-object trajectory consensus for
distributed fusion, and metrics for multi-object states and multi-object
trajectories. In Section \ref{sec:track-dist}, we formulate the Fr\'{e}chet
mean trajectory w.r.t. the OSPA trajectory metric, and propose an
efficient greedy algorithm to compute this trajectory mean. In Section
\ref{sec:mt-consensus} we introduce the OSPA-based\textsuperscript{}
metrics for measuring distance between multi-object trajectories and
the corresponding Fr\'{e}chet mean, together with a greedy method to efficiently
compute this mean. In Section \ref{sec:Gibbs-sampling}, we present
methods to compute trajectory and multi-object trajectory means based
on Gibbs sampling. Section \ref{sec:Experimental-Results} demonstrates
the performance of the proposed multi-object trajectory consensus
algorithm in a distributed multi-object tracking context. We conclude
the paper and suggest potential extensions in Section \ref{sec:Conclusion}.\vspace{-0.5em}

\section{Background\protect\label{sec:Background}}

This section discusses current techniques in distributed multi-object
tracking pertinent to multi-object trajectory averaging, and metrics
on multi-object states and trajectories that will be used to formulate
the mean trajectory and mean multi-object trajectory.\vspace{-0.9em}

\subsection{Distributed Multi-Object Estimation\protect\label{subsec:bckg-t2t}}

While there is no concept of mean or average multi-object trajectory,
researchers in distributed multi-object tracking have developed fusion
solutions based on indirect multi-object trajectory consensus. Assuming
multi-object densities--functions on the class of finite sets of
some (finite dimensional) space--are available, e.g., in the RFS
approach, a popular consensus technique is to take the average of
the relevant multi-object densities. The generalized covariance intersection
(GCI) approach \cite{mahler2000optimal}, also known as geometric
averaging (GA), has been used routinely to fuse multi-object densities.
GCI fusion algorithms for the probability hypothesis density (PHD)
filter \cite{mahler2003multitarget}, the cardinalized PHD filter
\cite{mahler2007phd} and the multi-Bernoulli (MB) filter \cite{vo2008cardinality}
have been proposed in \cite{ueney2013distributed}, \cite{battistelli2013consensus}
and \cite{wang2016distributed}, respectively. Additionally, GCI-based
fusion methods for the labeled multi-Bernoulli (LMB) filter \cite{reuter2014labeled}
and the generalized LMB (GLMB) filter \cite{vo2013labeled} have been
proposed in \cite{fantacci2018robust,li2018computationally,li2017robust,yi2019computationally}.
Alternatively, arithmetic averaging (AA) fusion has also been developed
for multi-object density fusion, including AA-PHD \cite{li2017clustering,li2019local},
AA-CPHD \cite{Gao2020multiobject}, AA-MB \cite{LiTian2020on}, and
AA-LMB/GLMB \cite{gao2020fusion}. The GA and AA fusion methods discussed
so far are intended for multi-object state fusion, and can be extended
to multi-object trajectory fusion via the multi-scan multi-object
density (i.e., joint density of the multi-object state over multiple
scans). However, this approach to fusing multi-object trajectories
is numerically infeasible due to the exponential growth in complexity. 

An alternative to multi-scan multi-object density fusion is track-to-track
(T2T) fusion \cite{chang1997optimal}. This is an intuitive technique
for fusing individual tracks together by minimizing some prescribed
cost functions. Additional information on the state can be augmented
into the cost function \cite{osbome2011track}. A popular cost function
is based on the association probability of the sets of estimated objects
with the physical objects \cite{bar1981track,ogle2019alternative}.
Alternatively, some authors use cost functions based on the statistical
distance/divergence between the distributions describing the states
of the tracks and their associated confidence \cite{tang2021multi,battistelli2014kullback,tang2018information},
or the OSPA distance between the tracks \cite{nguyen2021distributed}.
T2T assignment can be cast as a multi-dimensional assignment problem,
which is NP-hard for more than two dimensions. Nevertheless, sub-optimal
solutions can be obtained using efficient clustering techniques, including
hierarchical clustering \cite{bereson2006efficient}, DBSCAN \cite{he2018multi},
density-peak clustering \cite{chen2025distributed}, or stochastic
optimization \cite{wolf2023track}. In T2T fusion, clutter objects
are typically not considered within the optimization process but are
instead discarded based on existence probability \cite{muvsicki2015track},
voting logic \cite{zhang1997efficient}, or assignment confidence
\cite{stankovic2021adaptive}. Determining an effective cost function
for T2T fusion is still an active research topic \cite{wang2022consistent,kaplan2008assignment,bar1986effect}.

Numerically, multi-object trajectory fusion is more efficient with
T2T fusion since only the individual states or their distributions
are needed, rather than the entire multi-object densities. In \cite{bar2004multisensor},
the multi-scan assignment problem was solved using Lagrangian relaxation
for a small number of sensors. A fast algorithm for multi-scan T2T
assignment was proposed in \cite{tokta2018fast}, but it is limited
to two sensors. The OSPA-based multi-object trajectory fusion approach
proposed in \cite{nguyen2021distributed} has been shown to be effective
for a large number of sensors. Further, as a peer-to-peer fusion algorithm,
it is more robust to network disruption. However, it is less accurate
than global fusion algorithms, wherein all information is fused at
once.\vspace{-0.9em}

\subsection{Metrics for Point Patterns and Trajectories \protect\label{subsec:bckg-set-metrics}}

Since a multi-object state is a point pattern, metrics for point
patterns are natural for measuring the distance between multi-object
states. While the Hausdorff metric is an obvious point pattern metric,
it is insensitive to cardinality differences \cite{hoffman2004multitarget}.
To address this, the optimal mass transfer metric was proposed in
\cite{hoffman2004multitarget}, based on the Wasserstein construction.
However, this metric suffers from a physical inconsistency if the
cardinalities of the two multi-object states are not the same \cite{hoffman2004multitarget}.
A widely used multi-object metric with meaningful physical interpretations,
 capturing both the state and cardinality errors, is the OSPA metric
\cite{schuhmacher2008consistent}. The OSPA construction (by this
we mean the use of optimal sub-pattern assignment as the abbreviation
suggests) also admits other metrics such as the cardinalized optimal
linear assignment (COLA) metric \cite{barrios2017metrics}. The main
difference is that the COLA is not normalized by the cardinality.
Another unnormalized OSPA-based metric is a variant of the COLA metric,
misnamed as the Generalized OSPA (GOSPA) metric \cite{rahmathullah2017generalized},
which does not generalize the OSPA metric\footnote{GOSPA does not subsume the OSPA metric as a special case. In fact
GOSPA is still based on the OSPA construct. }\negthinspace{}. Indeed, it is equivalent to the transport-transform
\negthinspace{}(TT)\negthinspace{} metric proposed in \cite{muller2020metrics},
and thus, will be referred to as the TT metric since this is a more
appropriate name. Apart from the mathematical similarity, it has been
shown that COLA and TT metrics behave similarly in most scenarios,
but TT is more sensitive to variations in parameters \cite{barrios2023comparison}.

The TT and OSPA metrics are natural candidate distances to formulate
the Fr\'{e}chet mean of point patterns. Techniques for computing point
pattern barycentres w.r.t. the TT and OSPA metrics have been discussed
in \cite{muller2020metrics}. An earlier approach for computing the
OSPA barycenter using an alternating greedy search algorithm was proposed
in \cite{baum2015ospa}. This technique can be applied to approximate
the minimum mean OSPA estimate (from certain distributions), first
formulated in \cite{guerriero2010shooting} for fixed cardinality,
and later extended to unknown cardinality in \cite{balasingam2015mmospa}.

In many applications, each element of the point pattern represents
a probability distribution on $\mathbb{R}^{n}$ to describe the associated
uncertainty/confidence. Consequently, the discussed point pattern
metrics need to be extended accordingly. Recall that the constructions
of these metrics require a \textit{base metric} or \textit{base distance}
between the elements. If the elements belong to $\mathbb{R}^{n}$,
the $p$-norm of their difference is typically used as the base metric.
When the elements are probability distributions, metrics such as Wasserstein
\cite{assa2018wasserstein}, Hellinger, or geodesic distances on Gaussian
manifolds \cite{tang2018information,nielsen2023fisher} can be used
as the base metric. Indeed, the Hellinger metric was used as a base
metric in \cite{nagappa2011incorporating} to incorporate estimation
uncertainty into the OSPA metric. Further, information theoretic measures
such as Kullback-Leibler's divergence (KLD) \cite{battistelli2014kullback},
Jeffrey's divergence (i.e., symmetrized KLD) \cite{nielsen2009sided},
Bhattacharyya distance \cite{tang2021multi} can also be used to measure
the dissimilarity between distributions, though they may not satisfy
all metric properties.

In multi-object estimation, distances between multi-object trajectories
are equally as important as distances between multi-object states.
In \cite{ristic2011metric} and \cite{vu2014new}, extensions of the
OSPA metric (from multi-object states) to multi-object trajectories
were proposed. However, these metrics are intractable, or are no longer
proper metrics when numerically approximated in practice. A more practical
approach is to treat a multi-object trajectory as a point pattern
of trajectories, wherein point pattern metrics can be applied with
tractable base distances between trajectories. A number of trajectory
metrics were proposed in \cite{beard2017ospa}, but they exhibit counter-intuitive
behavior, except for the OSPA-based\textsuperscript{} construction.
The OSPA metric with the OSPA-based trajectory base metric, called
the OSPA\textsuperscript{(2)} metric, addresses the shortcomings
of previous multi-object trajectory metrics, offering meaningful physical
interpretation and can be efficiently computed for large-scale scenarios
\cite{beard2020solution}. The TGOSPA metric \cite{garcia2020ametric},
proposed after OSPA\textsuperscript{(2)}, is not applicable in this
work, because it was only proven for the special case of multi-object
trajectories with contiguous trajectories. When the trajectories are
sequences of probability distributions to describe the associated
uncertainty, the OSPA-based trajectory metric is constructed from
distances between probability distributions similar to that for the
OSPA metric for point patterns of distributions discussed above. Further
details and discussions on metrics for multi-object states and multi-object
trajectories can be found in \cite{nguyen2022trustworthy}. 

\section{Trajectory and Fr\'{e}chet Mean\protect\label{sec:track-dist}}

This section presents a formulation and an accompanying numerical
solution for trajectory consensus. Subsection \ref{subsec:OSPA-traj-dist}
introduces the OSPA trajectory distance and formulates the Fr\'{e}chet
mean trajectory w.r.t. this distance. A method to compute the Fr\'{e}chet
mean trajectory based on greedy search is presented in Subsection
\ref{subsec:comp-traj-mean}. A list of frequently used notation is
given in Table \ref{tab:List-of-notations}.\vspace{-0.9em}

\begin{table}[t]
\begin{onehalfspace}
\caption{List of frequently used notations.\protect\label{tab:List-of-notations}\vspace{-1em}}

\end{onehalfspace}

\centering{}%
\begin{tabular}{|c|l|}
\hline 
{\small\,\,\,}{\small\textbf{Notation}} & {\small\quad{}\quad{}\quad{}\quad{}\quad{}\,\,}{\small\textbf{Description}}\tabularnewline
\hline 
{\small$\mathbb{X}$} & {\small state space}\tabularnewline
{\small$\mathbb{K}$} & {\small finite time window $\{1,...,K\}$}\tabularnewline
{\small$\mathbb{T}$} & {\small space of trajectories}\tabularnewline
{\small$\mathcal{M}(\mathbb{T})$} & {\small space of multi-object trajectories}\tabularnewline
{\small$\mathbb{B}$} & {\small$\{0,1\}$}\tabularnewline
{\small$1:L$} & {\small$1,2,...,L$}\tabularnewline
{\small$\{1:L\}$} & {\small$\{1,2,...,L\}$}\tabularnewline
{\small$x_{1:L}$} & {\small$x_{1},x_{2},...,x_{L}$}\tabularnewline
{\small$x^{(1:L)}$} & {\small$x^{(1)},x^{(2)},...,x^{(L)}$}\tabularnewline
{\small$d_{\mathbb{X}}^{(c)}(\cdot,\cdot)$} & {\small a base distance on $\mathbb{X}$ with cut-off at $c$}\tabularnewline
{\small$d_{\mathbb{T}}^{(c)}(\cdot,\cdot)$} & {\small a trajectory distance on $\mathbb{T}$ with cut-off at $c$}\tabularnewline
{\small$\mathcal{D}_{\!u}$} & {\small domain of trajectory $u$}\tabularnewline
{\small$\mathcal{\bar{D}}_{\!u}$} & {\small complement of $\mathcal{D}_{\!u}$ in $\mathbb{K}$}\tabularnewline
{\small$\Pi_{L}$} & {\small the set of all permutation vectors on $\{1:L\}$}\tabularnewline
{\small$\mathbf{1}_{X}^{(x)}$ or $\mathbf{1}_{X}(x)$} & {\small equal 1 if $x\in X$, and equal 0 otherwise}\tabularnewline
\hline 
\end{tabular}
\end{table}

\subsection{The Fr\'{e}chet Mean Trajectory\protect\label{subsec:OSPA-traj-dist}}

Intuitively, the mean can be interpreted as the typical point of a
data set. In a vector space, the arithmetic mean (i.e., the sum of
the data points divided by the number of data points) minimizes the
variance (i.e., the total squared Euclidean distance) from the data
points. This property can be used to generalize the mean to an arbitrary
metric space\footnote{A space equipped with a metric for gauging the distance between any
two elements, and a metric is a non-negative function of two variables,
which satisfies the metric properties, see \cite{beard2020solution}.} where addition is not meaningful \cite{lou2020differentiating}.
More concisely, the notion of Fr\'{e}chet mean is given as follows.

\begin{definition}Let $(d,\mathbb{M})$ be a metric space and $r$
be a positive integer. For $v^{(1)},...,v^{(N)}\in\mathbb{M}$ and
weights $w^{(1)},...,w^{(N)}>0$, the $r^{th}$-order weighted \textit{Fr\'{e}chet
mean} is defined as
\begin{equation}
\hat{v}=\arg\min_{u\in\mathbb{M}}\sum_{n=1}^{N}w^{(n)}d^{r}(u,v^{(n)}).\label{eq:Fretchetmean}
\end{equation}
 When the weights are unity, $\hat{v}$ is called the \textit{Fr\'{e}chet
mean.}\end{definition}Note that the Fr\'{e}chet mean is not necessarily
unique in general. The above sum can be interpreted as the \textit{Fr\'{e}chet
variance} for $r=2$. On a vector space equipped with the Euclidean
distance, the above definition yields the arithmetic mean for $r=2$,
and the geometric median for $r=1$.

A trajectory is a time-sequence of states, and can be either contiguous
or fragmented \cite{beard2020solution,vo2024overview}. In a \textit{contiguous}
trajectory, the state exists at every instance from birth to death,
whereas in a \textit{fragmented} trajectory, there is at least an
instance between birth and death that the state does not exist. Fragmentation
manifests in the output of most tracking algorithms, mostly from incorrectly
declaring that the object ceases to exist due to misdetections. The
following definition covers both contiguous and fragmented trajectories
\cite{beard2020solution}.\vspace{-0em}

\begin{definition}A \textit{trajectory} $u$, on a state space $\mathbb{X}$
and a finite time window $\mathbb{K}$, is a function that maps $\mathbb{K}$
to $\mathbb{X}$. The set of instances (in $\mathbb{K}$) where the
state of trajectory $u$ exists is the domain of $u$, denoted as
$\mathcal{D}_{u}$. The space of all trajectories is denoted as $\mathbb{T}$.
\end{definition}\vspace{-0.5em}

Note that a trajectory $u$ is contiguous if $\mathcal{D}_{u}$ consists
of consecutive instances, otherwise it is fragmented. The state space
$\mathbb{X}$ is usually a (finite dimensional) vector space or the
space of probability distributions (on a vector space). The latter
is used to accommodate uncertainty on the states of the trajectory
at given instances. \vspace{-0em}

\begin{definition}\label{prop:ospa-track}Let $d_{\mathbb{X}}$ be
a metric on the state space $\mathbb{X}$, and $d_{\mathbb{X}}^{(c)}(y,z)\triangleq\min\{d_{\mathbb{X}}(y,z),c\}$,
$c>0$. For any trajectories $u$ and $v\in\mathbb{T}$, the OSPA
trajectory distance of order $r\in[1,\infty]$ and cut-off $c$ is
defined as \vspace{-0.5em}
\begin{multline}
\!\!\!\!\!d_{\mathbb{T}}^{(c,r)}(u,v)=\\
\!\!\!\left[\!\frac{{\scriptstyle {\displaystyle \sum_{k\in\mathcal{D}_{\!u}\cap\mathcal{D}_{\!v}}}\!}\!\!\![d_{\mathbb{X}}^{(c)}(u(k),v(k))]^{r}\!+\!c^{r}(|\mathcal{D}_{\!u}\mathcal{-D}_{\!v}|+|\mathcal{D}_{\!v}\mathcal{-D}_{\!u}|)}{|\mathcal{D}_{\!u}\cup\mathcal{D}_{\!v}|}\!\right]^{\!\frac{1}{r}}\!\!.\!\!\label{eq:ospa-track-metric}
\end{multline}
\end{definition}\vspace{-1em}\begin{remark}The proof of the OSPA
trajectory metric for $r=1$ was given in \cite{beard2020solution}.
Nonetheless, this result holds for any $r\in[1,\infty]$, and for
completeness see Supplementary Materials (SM) Subsection \ref{subsec:proof-prop:ospa-track}
for proof. When the states of both trajectories exist (i.e., $k\in\mathcal{D}_{\!u}\cap\mathcal{D}_{\!v}$)
the error is simply the distance between the respective states capped
by $c$, and when only one state exists the error is $c$. Thus, intuitively
the OSPA trajectory distance is simply the $r^{th}$-order time-averaged
state and cardinality errors between two trajectories.\end{remark} 

With respect to the metric $d_{\mathbb{T}}^{(c,r)}(\cdot,\cdot)$,
the $r^{th}$-order Fr\'{e}chet mean of the trajectories $v^{(1)},...,v^{(N)}\in\mathbb{T}$
(hereon abbreviated as $v^{(1:N)}$) is 
\begin{equation}
\hat{v}=\arg\min_{u\in\mathbb{T}}V^{(r)}(u),\label{eq:trackbary-cost}
\end{equation}
where
\begin{eqnarray}
V^{(r)}(u) & \!\!\!\!=\!\!\!\! & {\displaystyle \sum_{k\in\mathbb{K}}}\mathbf{1}_{\mathcal{D}_{u}}^{(k)}\Psi_{k}^{(r)}(u)+\mathbf{1}_{\mathcal{\overline{D}}_{u}}^{(k)}\bar{\Psi}_{k}^{(r)}(u),\\
\Psi_{k}^{(r)}(u) & \!\!\!\!=\!\!\!\! & {\displaystyle \sum_{n=1}^{N}}\frac{\mathbf{1}_{\mathcal{D}_{\!v^{(n)}\!}}^{(k)}[d_{\mathbb{X}}^{(c)\!}(u(k),v^{(n)\!}(k))]^{r}+\mathbf{1}_{\mathcal{\overline{D}}_{\!v^{(n)}\!}}^{(k)}c^{r}}{|\mathcal{D}_{u}\cup\mathcal{D}_{v^{(n)}}|},\\
\bar{\Psi}_{k}^{(r)}(u) & \!\!\!\!=\!\!\!\! & {\displaystyle \sum_{n=1}^{N}}\frac{\mathbf{1}_{\mathcal{D}_{\!v^{(n)}}}^{(k)}c^{r}}{|\mathcal{D}_{u}\cup\mathcal{D}_{\!v^{(n)}}|},
\end{eqnarray}
$\mathbf{1}_{X}^{(x)}=1$ if $x\in X$ and 0 otherwise, and $\mathcal{\overline{D}}_{u}=\mathbb{K}-\mathcal{D}_{u}$.
Note that the above functions implicitly depend on $v^{(1:N)}$.
The expression for the cost function $V^{(r)}(u)$ is obtained by
substituting (\ref{eq:ospa-track-metric}) into the sum in (\ref{eq:Fretchetmean})
with unit weights.

\begin{proposition}\label{prop:v-opt}Suppose $\hat{v}$ is an $r^{th}$-order
Fr\'{e}chet mean of the trajectories $v^{(1:N)}$. Then for each $k\in\mathcal{D}_{\hat{v}}$,
$\hat{v}(k)$ is the $r^{th}$-order Fr\'{e}chet mean of the trajectory
states $v^{(n)}(k)$ that exist at time $k$, weighted by $|\mathcal{D}_{\hat{v}}\cup\mathcal{D}_{v^{(n)}}|$,
i.e.,
\begin{eqnarray}
\hat{v}(k) & = & \arg\min_{\mu\in\mathbb{X}}{\displaystyle \sum_{n:\mathcal{D}_{\!v^{(n)}\!}\ni k}}\frac{[d_{\mathbb{X}}^{(c)\!}(\mu,v^{(n)\!}(k))]^{r}}{|\mathcal{D}_{\hat{v}}\cup\mathcal{D}_{v^{(n)}}|}.\label{eq:v_opt}
\end{eqnarray}
\end{proposition}The above result (see SM Subsection \ref{subsec:proof-prop:v-opt}
for proof) provides an explicit expression for the state of the mean
trajectory at each time step, given its domain. In the next subsection,
this result will be used to compute the mean trajectory.\vspace{-0em}

\subsection{Computing the OSPA Fr\'{e}chet Mean Trajectory\protect\label{subsec:comp-traj-mean}}

This subsection presents a method to compute the Fr\'{e}chet mean trajectory
by converting Problem (\ref{eq:trackbary-cost}) into an equivalent
form, and developing an efficient algorithm for solving it.

\subsubsection{Problem Transformation}

We transform the decision variables into an alternative representation
on a different space by decomposing a trajectory $u$ into its \textit{existence
history} $\gamma$ and \textit{states} $x$, via the transformation\vspace{-0em}
\begin{eqnarray*}
\mathcal{A}:\mathbb{T} & \rightarrow & \mathbb{B}^{|\mathbb{K}|}\times\mathbb{X}^{|\mathbb{K}|},\\
u & \mapsto & (\gamma,x),
\end{eqnarray*}
where $\mathbb{B}=\{0,1\}$, while the $k^{th}$ entries of $\gamma\in\mathbb{B}^{|\mathbb{K}|}$
and $x\in\mathbb{X}^{|\mathbb{K}|}$ are given by $\gamma_{k}=\mathbf{1}_{\mathcal{D}_{u}}^{(k)}$
and $x_{k}=\mathbf{1}_{\mathcal{D}_{u}}^{(k)}u(k)$, respectively
(here we use the convention $0\times u(k)=\boldsymbol{0}\in\mathbb{X}$
regardless of whether $u(k)$ is defined or not). If $\gamma_{k}=1$,
then the trajectory exists at the $k^{th}$ instance and its state
is given by $x_{k}$, otherwise, the trajectory does not exist at
the $k^{th}$ instance and $x_{k}$ is ignored. For example, the 1D
trajectory $u$, defined by $u(k)=k$ for each $k\in\{1,2,5\}$ on
the window $\mathbb{K}=\{1:5\},$ can be represented by $\gamma=[1,1,0,0,1]$
and $x=[1,2,0,0,5]$.

A trajectory can be recovered from its existence history and states,
by the transformation
\begin{eqnarray*}
\mathcal{R}:\mathbb{B}^{|\mathbb{K}|}\times\mathbb{X}^{|\mathbb{K}|} & \rightarrow & \mathbb{T},\\
(\gamma,x) & \mapsto & u,
\end{eqnarray*}
where $u(k)=x_{k}$ whenever $\gamma_{k}=1$, e.g., the 1D trajectory
$u$ with representation $\gamma=[1,1,0,0,1]$ and $x=[1,2,0,0,5]$,
can be reconstructed by setting $\mathcal{D}_{u}=\{1,2,5\}$, $u(1)=x_{1}=1$,
$u(2)=x_{2}=2$, and $u(5)=x_{5}=5$. It is clear that $\mathcal{R\circ A}=\mathbf{I}$,
where $\mathbf{I}$ is the identity function, but $\mathcal{R}$ is
not the inverse of $\mathcal{A}$ because there are more than one
element of $\mathbb{B}^{|\mathbb{K}|}\times\mathbb{X}^{|\mathbb{K}|}$
that $\mathcal{R}$ maps to the same $u$. 

The following lemma provides the sufficient conditions for the equivalence
of the original problem and its transformation.\vspace{0em}\begin{lemma}\label{lem:problem-transform-traj}Suppose
that the transformations $\mathcal{A}:\mathbb{Z}\rightarrow\mathbb{Y}$
and $\mathcal{R}:\mathbb{Y}\rightarrow\mathbb{Z}$ satisfy $\mathcal{R\circ A}=\mathbf{I}$
($\mathcal{A}$ is not necessarily invertible). Let $f:\mathbb{Z}\rightarrow[0,\infty)$
and $g:\mathbb{Y}\rightarrow[0,\infty)$ be such that
\begin{eqnarray*}
g\circ\mathcal{A}=f & \mathrm{and} & \mathcal{\mathcal{R}}(y)=\mathcal{\mathcal{R}}(y')\Rightarrow g(y)=g(y').
\end{eqnarray*}
 If $y^{*}$ minimizes $g$, then $\mathcal{R}(y^{*})$ minimizes
$f$, and $g(y^{*})=f(\mathcal{R}(y^{*}))$. (See SM Subsection \ref{subsec:proof-lem:problem-transform-traj}
for proof) \end{lemma}Using this equivalent representation, the
Fr\'{e}chet mean trajectory (\ref{eq:trackbary-cost}) is given by the
solution to Problem (\ref{eq:joint-gamma-x}) below on the space $\mathbb{B}^{|\mathbb{K}|}\times\mathbb{X}^{|\mathbb{K}|}$.
This is stated more concisely in the following (see SM Subsection
\ref{subsec:proof-prop:transformed-problem-traj} for proof). 

\begin{proposition}\label{prop:transformed-problem-traj}An $r^{th}$-order
Fr\'{e}chet mean of the trajectories $v^{(1:N)}$ is the trajectory $\hat{v}$,
defined by $\hat{v}(k)=\hat{x}_{k}$ whenever $\hat{\gamma}_{k}=1$,
where 
\begin{eqnarray}
(\hat{\gamma},\hat{x}) & \!\!\!\!=\!\!\!\! & \arg\min_{(\gamma,x)\in\mathbb{B}^{|\mathbb{K}|}\times\mathbb{X}^{|\mathbb{K}|}}U^{(r)}(\gamma,x),\label{eq:joint-gamma-x}\\
U^{(r)}(\gamma,x) & \!\!\!\!=\!\!\!\! & {\displaystyle \sum_{k\in\mathbb{K}}}\gamma_{k}\psi_{k}^{(r)}(\gamma,x_{k})+\left(1-\gamma_{k}\right)\bar{\psi}_{k}^{(r)}(\gamma),\label{eq:Ur}\\
\!\!\!\psi_{k}^{(r)}(\gamma,x_{k}) & \!\!\!\!=\!\!\!\! & {\displaystyle \sum_{n=1}^{N}\!}\frac{\mathbf{1}_{\mathcal{D}_{\!v^{(n)}\!}}^{(k)}[d_{\mathbb{X}}^{(c)\!}(x_{k},v^{(n)\!}(k))]^{r}\!+\!\mathbf{1}_{\mathcal{\overline{D}}_{\!v^{(n)}\!}}^{(k)}c^{r}}{|\mathcal{D}_{\gamma}\cup\mathcal{D}_{v^{(n)}}|},\\
\bar{\psi}_{k}^{(r)}(\gamma) & \!\!\!\!=\!\!\!\! & {\displaystyle \sum_{n=1}^{N}}\frac{\mathbf{1}_{\mathcal{D}_{\!v^{(n)}}}^{(k)}c^{r}}{|\mathcal{D}_{\gamma}\cup\mathcal{D}_{\!v^{(n)}}|},\\
\mathcal{D}_{\gamma} & \!\!\!\!=\!\!\!\! & \{k\in\mathbb{K}:\gamma_{k}=1\}.\label{eq:Ur-2}
\end{eqnarray}
\vspace{-1em}\end{proposition}Further, Problem (\ref{eq:joint-gamma-x})
can be reduced to optimizing on $\mathbb{B}^{|\mathbb{K}|}$ by exploiting
Proposition \ref{prop:v-opt} as follows. For each $\gamma\in\mathbb{B}^{|\mathbb{K}|}$,
let
\begin{eqnarray}
x^{*}(\gamma) & \!\!\!\!\triangleq\!\!\!\! & \arg\min_{x\in\mathbb{X}^{|\mathbb{K}|}}U^{(r)}(\gamma,x),\label{eq:joint-gamma-x-1}\\
W^{(r)}(\gamma) & \!\!\!\!\triangleq\!\!\!\! & U^{(r)}(\gamma,x^{*}(\gamma)).\label{eq:Ur-1}
\end{eqnarray}
It is clear that if $\gamma^{*}$ minimizes $W^{(r)}$, then $(\gamma^{*},x^{*}(\gamma^{*}))$
minimizes $U^{(r)}$, because $U^{(r)}(\gamma^{*},x^{*}(\gamma^{*}))=W^{(r)}(\gamma^{*})$,
and $W^{(r)}(\gamma^{*})\leq U^{(r)}(\gamma,x^{*}(\gamma))\leq U^{(r)}(\gamma,x)$.
Using the property of the Fr\'{e}chet mean in Proposition \ref{prop:v-opt}
we can compute $x^{*}(\gamma)$, and subsequently $W^{(r)}(\gamma)$
for any $\gamma$ as summarized in Proposition \ref{prop:mean-single-state}
below. \textit{Hence, solving Problem (\ref{eq:joint-gamma-x}) amounts
to finding the minimizer $\gamma^{*}$ of $W^{(r)}$, and then invoking
Proposition \ref{prop:mean-single-state} again to determine the states
of the mean trajectory.} 

\begin{proposition}\label{prop:mean-single-state}

For any $\gamma\in\mathbb{B}^{|\mathbb{K}|}$, let $[x_{1}^{*}(\gamma),...,x_{|\mathbb{K}|}^{*}(\gamma)]\triangleq\arg\min_{x\in\mathbb{X}^{|\mathbb{K}|}}U^{(r)}(\gamma,x)$.
If $\gamma_{k}\!=\!0$ or $\ \mathbf{1}_{\mathcal{D}_{\!v^{(n)}\!}}^{(k)}=\!0$,
for all $n=\!1:N$, then $x_{k}^{*}(\gamma)\!=\!\boldsymbol{0}\in\mathbb{X}$.
Further, if $\gamma_{k}\!=\!1$, then 
\begin{equation}
x_{k}^{*}(\gamma)=\arg\min_{x_{k}\in\mathbb{X}}{\displaystyle \sum_{n:\mathcal{D}_{\!v^{(n)}\!}\ni k}}\frac{[d_{\mathbb{X}}^{(c)\!}(x_{k},v^{(n)\!}(k))]^{r}}{|\mathcal{D}_{\gamma}\cup\mathcal{D}_{v^{(n)}}|}.\label{eq:xk-star}
\end{equation}
\end{proposition}This result (see SM Subsection \ref{subsec:proof-prop:mean-single-state}
for proof) provides an explicit expression for each component of $x^{*}(\gamma)$.
Specifically, if at time $k$, the state of $x^{*}(\gamma)$ does
not exist ($\gamma_{k}\!=\!0$) or all states of the sample trajectories
do not exist, then $x_{k}^{*}(\gamma)$ is set to $\boldsymbol{0}\in\mathbb{X}$.
If the state of $x^{*}(\gamma)$ exists at time $k$ ($\gamma_{k}\!=\!1$),
then $x_{k}^{*}(\gamma)$ is given by the $r^{th}$-order Fr\'{e}chet
mean of the sample trajectory states that exist, weighted by $|\mathcal{D}_{\gamma}\cup\mathcal{D}_{v^{(n)}}|$.
Note that $x_{k}^{*}(\gamma)$ in (\ref{eq:xk-star}) may not be unique,
but any minimizer of the cost would suffice.

\subsubsection{Mean Trajectory via Greedy Search\protect\label{subsec:Greedy-Algorithm-Trajectory}}

Minimizing $W^{(r)}$ on $\mathbb{B}^{|\mathbb{K}|}$ is a discrete
optimization problem that can be solved with non-linear binary programming
or other stochastic optimization techniques such as simulated annealing
or genetic algorithms \cite{kochenderfer2019algorithms}. Exhaustive
search is also possible if the search space $\mathbb{B}^{|\mathbb{K}|}$
is small ($|\mathbb{K}|$ is small). 

Targeting efficiency to suit online applications, we present a greedy
search algorithm that generates a new $\tilde{\gamma}$ from $\gamma$
simply by changing the component $\gamma_{k}\in\{0,1\}$ to its complement
(i.e., from 0 to 1 or vice-versa) if this reduces the cost $W^{(r)}$.
This process is cycled through all values of $k\in\mathbb{K}$ until
the cost reduction falls below some threshold. The pseudocode is given
in Algorithm \ref{alg:greedy-existence}. If the complexity of solving
Problem (\ref{eq:xk-star}) is $\mathcal{O}(M)$, then the worst case
complexity of this algorithm is $\mathcal{O}(T_{t}K^{2}M)$. Figure
\ref{fig:samples-mean-so} shows the mean trajectories computed using
this method. 
\begin{algorithm}
\textbf{Output}: Mean trajectory $\hat{v}$.\vspace{-0.5em}

\rule[0.5ex]{1\columnwidth}{1pt}

Initialize $\gamma^{*}:=\boldsymbol{0}$;

Set $\gamma_{k}^{*}:=1$ if $\sum_{n=1}^{N}\mathbf{1}_{\mathcal{D}_{\!v^{(n)}}}^{(k)}\!>\!0$;

$\textrm{cost}^{*}:=$ $W^{(r)}(\gamma^{*})$; $\delta_{\textrm{cost}}:=\infty$;
$i:=0$;

\textbf{while} $i<T_{t}$ \textbf{and} $\delta_{\textrm{cost}}>\epsilon_{t}$
\textbf{do}

\quad{}$i$++;

\quad{}\textbf{for} $k:=1:K$ \textbf{do }$\tilde{\gamma}:=\gamma^{*}$;
$\tilde{\gamma}_{k}:=1-\gamma_{k}^{*}$;

\quad{}\quad{}\textbf{if} $\textrm{\ensuremath{W^{(r)}}(\ensuremath{\tilde{\gamma}})}<\textrm{cost}^{*}$
\textbf{then}

\quad{}\quad{}\quad{}$\gamma^{*}:=\tilde{\gamma}$; $\delta_{\textrm{cost}}\!:=\!$
$\textrm{cost}^{*}-W^{(r)}(\tilde{\gamma})$; $\textrm{cost}^{*}:=W^{(r)}(\tilde{\gamma})$;

Compute $x^{*}$ from $\gamma^{*}$ using (\ref{eq:xk-star});

Transform $\hat{v}:=\mathcal{R}(\gamma^{*},x^{*})$;

\caption{Greedy algorithm for trajectory consensus with $T_{t}$ iterations
and convergence threshold $\epsilon_{t}$.\protect\label{alg:greedy-existence}}
\end{algorithm}

Observe from (\ref{eq:Ur})-(\ref{eq:joint-gamma-x-1}) that the dependence
of trajectory $x^{*}(\gamma)$ on $\gamma$ is captured in the factors
$\left(|\mathcal{D}_{\gamma}\cup\mathcal{D}_{v^{(n)}}|\right)^{-1}$,
$n\in\{1:N\}$. In practice, these factors have similar values for
different $n$ since the estimates of the same trajectory, but by
different nodes, would have similar lengths. Moreover, these lengths
are relatively long, thereby diminishing the influence of $\left(|\mathcal{D}_{\gamma}\cup\mathcal{D}_{v^{(n)}}|\right)^{-1}$
on $x^{*}(\gamma)$. Hence, to further reduce computations, the optimal
$x^{*}(\gamma)$ can be approximated by an $x^{*}$ with components
\begin{equation}
x_{k}^{*}=\arg\min_{x_{k}\in\mathbb{X}}\sum_{n:\mathcal{D}_{\!v^{(n)}\!}\ni k}[d_{\mathbb{X}}^{(c)}(x_{k},v_{k}^{(n)})]^{r}.\label{eq:approx-x}
\end{equation}
If $\{1:N\}\cap\{n:\mathcal{D}_{\!v^{(n)}\!}\ni k\}=\emptyset$, we
set $x_{k}=\boldsymbol{0}\in\mathbb{X}$. This approximation reduces
the worst case complexity of Algorithm \ref{alg:greedy-existence}
to $\mathcal{O}(K(T_{t}+M))$. For the examples shown in Figure \ref{fig:samples-mean-so},
this approximate cost produces almost identical mean trajectories
to those from the exact cost.

\section{Multi-Object Trajectory and Fr\'{e}chet Mean\protect\label{sec:mt-consensus}}

This section extends the notion of consensus for trajectories in the
preceding section to multi-object trajectories. We start by formulating
the Fr\'{e}chet mean multi-object trajectory w.r.t. suitable OSPA-based
metrics in Subsection \ref{subsec:mt-consensus-distance}, and then
transform this problem to an equivalent form amenable to tractable
solutions in Subsection \ref{subsec:mt-consensus-formulation}.\vspace{-1em}

\subsection{The Fr\'{e}chet Mean Multi-Object Trajectory\protect\label{subsec:mt-consensus-distance}}

A multi-object trajectory $\boldsymbol{X}$ is a point pattern\footnote{A point pattern is a multi-set and may contain repeated elements.}
of trajectories, i.e., each element of $\boldsymbol{X}$ belongs to
the space $\mathbb{T}$ of trajectories. Let $\mathcal{M}(\mathbb{T})$
denote the space of all multi-object trajectories. Then the $r^{th}$-order
Fr\'{e}chet mean of the multi-object trajectories $\boldsymbol{Y}^{(1:N)}=\boldsymbol{Y}^{(1)},...,\boldsymbol{Y}^{(N)}$,
w.r.t. a multi-object trajectory metric $\boldsymbol{d}$, is given
by
\begin{equation}
\hat{\boldsymbol{Y}}=\arg\min_{\boldsymbol{X}\in\mathcal{M}(\mathbb{T})}\sum_{n=1}^{N}\boldsymbol{d}^{r}(\boldsymbol{X},\boldsymbol{Y}^{(n)}).\label{eq:mean-multi-trajectory-def}
\end{equation}
Various multi-object trajectory metrics can be constructed using the
OSPA-trajectory metric (presented in Section \ref{subsec:OSPA-traj-dist})
as the base metric, e.g., the Hausdorff-OSPA, Wasserstein-OSPA and
OSPA-OSPA or OSPA\textsuperscript{(2)} metric \cite{beard2020solution,nguyen2022trustworthy}.
The Hausdorff is insensitive to cardinality differences \cite{hoffman2004multitarget}
while the Wasserstein suffers from a physical inconsistency in cardinalities
\cite{hoffman2004multitarget}. The OSPA\textsuperscript{(2)} metric,
intuitively capturing the time-averaged cardinality and state errors
per trajectory, alleviates these issues and is defined as follows. 

\begin{definition}Consider $r\geq1$, $p\geq0$, $c\in[p,\sqrt[r]{2}p]$,
and the OSPA trajectory distance $d_{\mathbb{T}}^{(c,r)}$ on $\mathbb{T}$
given by\footnote{We can use $d_{\mathbb{T}}^{(c,r)}(y,z)=\min\{d_{\mathbb{T}}^{(q,l)}(y,z),c\}$,
where $d_{\mathbb{T}}^{(q,l)}(y,z)$ is the OSPA trajectory distance,
but for convenience we set $q=c$ and $l=r$.} (\ref{eq:ospa-track-metric}). For any two multi-object trajectories
$\boldsymbol{X}=\{\boldsymbol{X}_{1:m}\}$ and $\boldsymbol{Y}=\{\boldsymbol{Y}_{1:s}\}$
(without loss of generality, assume $m\leq s$), the \textit{OSPA}\textsuperscript{(2)}\textit{
}(\textit{multi-object trajectory})\textit{ distance} of \textit{order}
$r$, \textit{cardinality penalty} $p$ and\textit{ cut-off} $c$
is defined as \cite{muller2020metrics,schuhmacher2008consistent,beard2020solution}
\begin{equation}
\!\!\!\boldsymbol{d}(\boldsymbol{X}\!,\boldsymbol{Y})=\left[\min_{\pi\in\Pi_{s}}\!\sum_{\ell=1}^{m}\!\frac{[d_{\mathbb{T}}^{(c,r)}\!(\boldsymbol{X}_{\!\ell},\boldsymbol{Y}_{\!\!\pi(\ell)})]^{r}+|s\!-\!m|p^{r}}{s}\!\right]^{\!\frac{1}{r}}\!\!,\label{eq:ospa^(2)}
\end{equation}
with $\boldsymbol{d}(\emptyset\!,\emptyset)=0$, where $\Pi_{s}$
denotes the space of permutations on $\{1:s\}$. If $m>s$, we define
$\boldsymbol{d}(\boldsymbol{X}\!,\boldsymbol{Y})=\boldsymbol{d}(\boldsymbol{Y}\!,\boldsymbol{X})$.
Note that for simplicity, we suppressed the dependence on the parameters
$r,$ $c$ and $p$. This definition is also based on the fact that
the OSPA metric \cite{schuhmacher2008consistent} can be extended
to the case $c\in[p,\sqrt[r]{2}p]$, termed the relative TT metric
in \cite{muller2020metrics} (see Section 2 of the referred paper
for more details).\end{definition}\begin{remark}If $s=m$, (\ref{eq:ospa^(2)})
is also the Wasserstein distance.\end{remark}

More generally $\boldsymbol{d}(\boldsymbol{X}\!,\boldsymbol{Y})$
is the OSPA distance between the multi-object trajectories $\boldsymbol{X}\!,\boldsymbol{Y}$
when $d_{\mathbb{T}}^{(c)}(y,z)=\min\{d_{\mathbb{T}}(y,z),c\}$, for
any trajectory distance $d_{\mathbb{T}}(y,z)$. Note also that the
OSPA construction admits variations of the OSPA metric such as the
COLA metric \cite{barrios2017metrics}, and the TT metric \cite{muller2020metrics}
by extending (\ref{eq:ospa^(2)}) to 
\begin{eqnarray}
\!\!\!\!\boldsymbol{d}(\boldsymbol{X}\!,\boldsymbol{Y})=\left[\min_{\pi\in\Pi_{s}}\!\sum_{\ell=1}^{m}\!\frac{[d_{\mathbb{T}}^{(c)}\!(\boldsymbol{X}_{\!\ell},\boldsymbol{Y}_{\!\!\pi(\ell)})]^{r}+|s\!-\!m|p^{r}}{\kappa(s,m)}\right]^{\!\frac{1}{r}}\!\!, & \!\label{eq:ospa-based-traj-ori}
\end{eqnarray}
where $\kappa$ is a non-negative real function of non-negative integer
pairs such that $\kappa(s,m)=\kappa(m,s)$. Setting: $\kappa(s,m)=\max(s,m)$
yields OSPA\textsuperscript{(2)}; $\kappa(s,m)=c^{r}$, $p=c$ yields
COLA-OSPA; and $\kappa(s,m)=1$, $p=c/\sqrt[r]{\alpha}$ yields TT-OSPA.\vspace{1.5em}

An upper bound on the cardinality of the Fr\'{e}chet mean multi-object
trajectory w.r.t. the OSPA-based metrics is given in the following
proposition. The proof is given in SM Subsection \ref{subsec:proof-prop-max-ospa-mean-card}.
\begin{proposition}\label{prop:max-ospa-mean-card}Consider the Fr\'{e}chet
mean of $\boldsymbol{Y}^{(1:N)}$ w.r.t. the OSPA-based metric described
by (\ref{eq:ospa-based-traj-ori}). If for each $j\geq L\triangleq\sum_{n=1}^{N}|\boldsymbol{Y}^{(n)}|$,
$\kappa(j,|\boldsymbol{Y}^{(n)}|)$ depends only on $j$, i.e. $\kappa(j,|\boldsymbol{Y}^{(n)}|)=\kappa(j)$,
and $\kappa(j+1)-\kappa(j)\leq\kappa(L)/L$. Then, the Fr\'{e}chet mean's
cardinality is no greater than $L$.\end{proposition} Note that OSPA\textsuperscript{(2)},
COLA-OSPA and TT-OSPA all satisfy the premise of the above proposition.
The next subsection presents a tractable method to compute the Fr\'{e}chet
mean in (\ref{eq:mean-multi-trajectory-def}) w.r.t. an OSPA-based
multi-object trajectory metric of the form (\ref{eq:ospa-based-traj-ori}).\vspace{-1em}

\subsection{Computing the Fr\'{e}chet Mean Multi-Object Trajectory\protect\label{subsec:mt-consensus-formulation}}

Computing the mean multi-object trajectory requires solving an optimization
problem on the space $\mathcal{M}(\mathbb{T})$, which is difficult
since the decision variable involves both the cardinality and elements
of the multi-object trajectory. In this subsection, we convert this
problem into an equivalent form, and present an efficient algorithm
for solving it.

\subsubsection{Problem Transformation\protect\label{subsec:mt-consensus-cost-function}}

Similar to Subsection \ref{subsec:comp-traj-mean}, we transform the
decision variable in $\mathcal{M}(\mathbb{T})$ to an alternative
representation on a different space by decomposing a multi-object
trajectory $\boldsymbol{X}=\{\boldsymbol{X}_{1:m}\}$ into the \textit{existence
list} $\eta$ and \textit{trajectory list} $\tau$, each with length
$L\geq m$, via the transformation
\begin{eqnarray*}
\mathcal{A}:\mathcal{M}(\mathbb{T}) & \rightarrow & \mathbb{B}^{L}\times\mathbb{T}^{L},\\
\{\boldsymbol{X}_{1:m}\} & \mapsto & (\eta,\tau),
\end{eqnarray*}
where the list entries $\eta_{\ell}=1$, $\tau_{\ell}=\boldsymbol{X}_{\ell}$
for $\ell=1:m$, and $\eta_{\ell}=0$, $\tau_{\ell}=\boldsymbol{0}$
(defined to be function that maps every element of $\mathbb{K}$ to
zero), otherwise. By convention, $\boldsymbol{X}=\{\}$ is represented
by existence and trajectory lists of all zeros. 

If $\eta_{\ell}=1$, then the $\ell^{th}$ trajectory exists and takes
on value $\tau_{\ell}$, otherwise the trajectory does not exist and
whatever value of $\tau_{\ell}$ is ignored. A multi-object trajectory
can be recovered from its existence and trajectory lists by the transformation
\begin{eqnarray*}
\mathcal{R}:\mathbb{B}^{L}\times\mathbb{T}^{L} & \rightarrow & \mathcal{M}(\mathbb{T}),\\
(\eta,\tau) & \mapsto & \boldsymbol{X}=\{\tau_{\ell}:\eta_{\ell}=1\},
\end{eqnarray*}
It is clear that $\mathcal{R\circ A}=\mathbf{I}$, but $\mathcal{R}$
is not the inverse of $\mathcal{A}$ because there are more than one
element of $\mathbb{B}^{L}\times\mathbb{T}^{L}$ that $\mathcal{R}$
maps to the same $\boldsymbol{X}$. 

Since we are interested in computing the mean of the multi-object
trajectories $\boldsymbol{Y}^{(1:N)}$, the list length $L$ should
not be less than the cardinality of this mean. Using the upper bound
on the Fr\'{e}chet mean in Proposition \ref{prop:max-ospa-mean-card},
we fix $L=\sum_{n=1}^{N}|\boldsymbol{Y}^{(n)}|$.

Note that Problem (\ref{eq:mean-multi-trajectory-def})-(\ref{eq:ospa-based-traj-ori})
requires computing an OSPA-based metric, which involves minimizing
over the permutation space parameterized by the cardinality of the
decision variable in $\mathcal{M}(\mathbb{T})$. Mapping the decision
variable into $\mathbb{B}^{L}\times\mathbb{T}^{L}$ results in transforming
this permutation space to the space of existence assignments defined
as follows. 

\begin{definition}Given the existence lists $\eta$, $\xi^{(1:N)}$
in $\mathbb{B}^{L}$, an \textit{existence assignment} between $\eta$
and $\xi^{(n)}$ is a permutation $\pi$ of $\{1:L\}$ such that:
\begin{itemize}
\item $\eta_{\ell}=1\Rightarrow\xi_{\pi(\ell)}^{(n)}=1$, if $\left\Vert \eta\right\Vert _{1}\leq\left\Vert \xi^{(n)}\right\Vert _{1}$;
or 
\item $\xi_{\pi(\ell)}^{(n)}=1\Rightarrow\eta_{\ell}=1$, if $\left\Vert \eta\right\Vert _{1}>\left\Vert \xi^{(n)}\right\Vert _{1}$.
\end{itemize}
\end{definition}

We denote $\Omega^{(n)}(\eta)$ as the space of existence assignments
between $\eta$ and $\xi^{(n)}$, and $\Omega^{(1:N)}(\eta)\triangleq\Omega^{(1)}(\eta)\times...\times\Omega^{(N)}(\eta)$.
Any element of $\Omega^{(n)}(\eta)$ can be represented as an $L$-vector
$\omega^{(n)}=[\omega^{(n)}(\ell_{1}),...,\omega^{(n)}(\ell_{L})]$,
and any $\omega\in\Omega^{(1:N)}(\eta)$, can be represented as an
$N\times L$ matrix with $(n,\ell)$ entry $\omega(n,\ell)=\omega^{(n)}(\ell)$.

Using the transformation $\mathcal{A}$ above, Problem (\ref{eq:mean-multi-trajectory-def})-(\ref{eq:ospa-based-traj-ori})
on $\mathcal{M}(\mathbb{T})$ is equivalent to Problem (\ref{eq:multi-traj-mean-problem-trans})
on $\mathbb{B}^{L}\times\mathbb{T}^{L}\times\left[\Pi_{L}\right]^{N}$.
This is stated more concisely in the following (see SM Subsection
\ref{subsec:proof-prop:set-trajectory-problem-transform} for proof). 

\begin{proposition}\label{prop:set-trajectory-problem-transform}Let
$\xi^{(1:N)}$, $\chi^{(1:N)}$ denote the respective existence and
trajectory lists of the multi-object trajectories $\boldsymbol{Y}^{(1:N)}$,
and define the functions
\begin{align}
Q_{\omega}^{(r)}(\eta,\tau) & =\sum_{\ell=1}^{L}\eta_{\ell}\phi_{\omega,\ell}^{(r)}(\eta,\tau_{\ell})+(1-\eta_{\ell})\bar{\phi}_{\omega,\ell}^{(r)}(\eta),\label{eq:cost-multi-traj-mean-problem-trans}\\
\phi_{\omega,\ell}^{(r)}(\eta,\tau_{\ell}) & =\sum_{n=1}^{N}\!\frac{\xi_{\omega(n,\ell)}^{(n)}[d_{\mathbb{T}}^{(c)\!}(\tau_{\ell},\chi_{\omega(n,\ell)}^{(n)})]^{r}+(1\!-\!\xi_{\omega(n,\ell)}^{(n)})p^{r}}{\kappa(|\!|\eta|\!|_{1},|\!|\xi^{(n)}|\!|_{1})},\\
\bar{\phi}_{\omega,\ell}^{(r)}(\eta) & =\sum_{n=1}^{N}\!\frac{\xi_{\omega(n,\ell)}^{(n)}p^{r}}{\kappa(|\!|\eta|\!|_{1},|\!|\xi^{(n)}|\!|_{1})},
\end{align}
(which implicitly depend on $\xi^{(1:N)}$ and $\chi^{(1:N)}$). If
\begin{align}
(\hat{\eta},\hat{\tau}) & =\arg\min_{(\eta,\tau)\in\mathbb{B}^{L}\times\mathbb{T}^{L}}\min_{\omega\in\Omega^{(1:N)}(\eta)}Q_{\omega}^{(r)}(\eta,\tau),\label{eq:multi-traj-mean-problem-trans}
\end{align}
then the multi-object trajectory $\boldsymbol{\hat{X}}=\{\hat{\tau}_{\ell}:\hat{\eta}_{\ell}=1\}$
is an $r^{th}$-order Fr\'{e}chet mean of $\boldsymbol{Y}^{(1:N)}$.\end{proposition}Further,
Problem (\ref{eq:multi-traj-mean-problem-trans}) reduces to optimizing
on $\mathbb{B}^{L}\times\left[\Pi_{L}\right]^{N}$ as follows. For
each $\eta\in\mathbb{B}^{L}$ and $\omega\in\left[\Pi_{L}\right]^{N}$,
let 
\begin{align}
\tau^{*}(\eta,\omega) & \triangleq\arg\min_{\tau\in\mathbb{T}^{L}}Q_{\omega}^{(r)}(\eta,\tau),\\
S^{(r)}(\eta,\omega) & \triangleq Q_{\omega}^{(r)}(\eta,\tau^{*}(\eta,\omega)),
\end{align}
and restricting ourselves to the constraint $\omega\in\Omega^{(1:N)}(\eta)$,
i.e., 
\begin{equation}
\boldsymbol{1}_{\Omega^{(n)}(\eta_{1:L})}(\omega^{(n)})=1,n\in\{1:N\}.\label{eq:constraint}
\end{equation}
If $(\eta^{*},\omega^{*})$ minimizes $S^{(r)}$, subject to constraint
(\ref{eq:constraint}), then $(\eta^{*},\tau^{*}(\eta^{*},\omega^{*}))$
minimizes $Q_{\omega^{*}}^{(r)}$, because $Q_{\omega^{*}}^{(r)}(\eta^{*},\tau^{*}(\eta^{*},\omega^{*}))=S^{(r)}(\eta^{*},\omega^{*})$
and $S^{(r)}(\eta^{*},\omega^{*})\leq Q_{\omega^{*}}^{(r)}(\eta,\tau^{*}(\eta,\omega^{*}))\leq Q_{\omega^{*}}^{(r)}(\eta,\tau)$.
In addition, Proposition \ref{prop:mean-single-trajectory} below
enables the computation of $\tau^{*}(\eta,\omega)$, and subsequently
$S^{(r)}(\eta,\omega)$, for any feasible $(\eta,\omega)$. \textit{Hence,
Problem (\ref{eq:multi-traj-mean-problem-trans}) translates to finding
the minimizer $(\eta^{*},\omega^{*})$ of $S^{(r)}$, and invoking
Proposition \ref{prop:mean-single-trajectory} again to determine
the constituent trajectories of the optimal solution.}\vspace{1em}

\begin{proposition}\label{prop:mean-single-trajectory}For any $\eta\!\in\!\mathbb{B}^{L}$
and $\omega\!\in\!\Omega^{(1:N)}(\eta)$, let $[\tau_{1}^{*}(\eta,\omega),...,\tau_{L}^{*}(\eta,\omega)]=\arg\min_{\tau\in\mathbb{T}^{L}}Q_{\omega}^{(r)}(\eta,\tau)$.
If $\eta_{\ell}=0$, or $\xi_{\omega(n,\ell)}^{(n)}=0$ for all $n=\!1\!:\!N$,
then $\tau_{\ell}^{*}(\eta,\omega)\!=\!\boldsymbol{0}$. Otherwise,
\begin{equation}
\tau_{\ell}^{*}(\eta,\omega)\!=\!\arg\min_{\tau_{\ell}\in\mathbb{T}}\!\sum_{n:\xi_{\omega(n,\ell)}^{(n)}=1}\!\!\frac{[d_{\mathbb{T}}^{(c)}\!(\tau_{\ell},\chi_{\omega(n,\ell)}^{(n)})]^{r}}{\kappa(|\!|\eta|\!|_{1},|\!|\xi^{(n)}|\!|_{1})}.\label{eq:mean_each_traj}
\end{equation}
\end{proposition}This result (see SM Subsection \ref{subsec:proof-prop:set-trajectory-problem-transform-2}
for proof) provides an explicit expression for each component trajectory
$\tau_{\ell}^{*}(\eta,\omega)$ of the trajectory list $\tau^{*}(\eta,\omega)$
that minimizes $Q_{\omega}^{(r)}(\eta,\cdot)$. Specifically, the
$\ell^{th}$ trajectory $\tau_{\ell}^{*}(\eta,\omega)$ is set to
$\boldsymbol{0}\in\mathbb{T}$, if $\eta_{\ell}=0$ or $\tau_{\ell}^{*}(\eta,\omega)$
is not associated (via $\omega$) with any sample trajectories. Otherwise,
$\tau_{\ell}^{*}(\eta,\omega)$ is the $r^{th}$-order Fr\'{e}chet mean
of the associated sample trajectories $\chi_{\omega(1,\ell)}^{(1)}$,
..., $\chi_{\omega(N,\ell)}^{(N)}$ that exist, weighted by $\kappa(|\!|\eta|\!|_{1},|\!|\xi^{(1)}|\!|_{1})$,
..., $\kappa(|\!|\eta|\!|_{1},|\!|\xi^{(N)}|\!|_{1})$, which can
be computed using the method in Subsection \ref{subsec:Greedy-Algorithm-Trajectory}.
Note that the solution of Problem (\ref{eq:mean_each_traj}) might
not be unique. Nonetheless, any minimizer of the cost would be sufficient.

\subsubsection{Mean Multi-Object Trajectory via Greedy Search\protect\label{subsec:mt-greedy}}

Minimizing the cost function $S^{(r)}\!$ on $\mathbb{B}^{L}\!\times\!\left[\Pi_{L}\right]\!$
subject to constraint (\ref{eq:constraint}) is generally intractable.
Inspired by the optimization method in \cite{muller2020metrics},
we present a greedy algorithm that alternates between $\eta$ and
$\omega$ until some local convergence condition is satisfied. This
is outlined in Algorithm \ref{alg:bary-ospa}. First, we initialize
$\eta\!=\!\boldsymbol{0}$ and any existence assignment matrix $\omega\!\in\!\Omega^{(1:N)}\!(\eta)$.
We then alternate between minimizing: \textit{i)} on the space $\Omega^{(1:N)}\!(\eta)$
to update $\omega$; and \textit{ii)} on the space $\mathbb{B}^{L}\!$
to update\texttt{ $\!\eta$}. 
\begin{algorithm}
\textbf{Output}: Mean multi-object trajectory $\hat{\boldsymbol{X}}$.\vspace{-0.5em}

\rule[0.5ex]{1\columnwidth}{1pt}

Initialize $\eta:=\boldsymbol{0}$, $\omega\in\Omega^{(1:N)}(\eta)$;

$\omega,\textrm{cost}^{*}$ := $\mathtt{optimizePermutation}$$(\eta,\omega)$;

$\eta^{*}:=\eta$; $\omega^{*}$:= $\omega$; $\delta_{\textrm{cost}}:=\infty$;
$i:=0$;

\textbf{while} $i<$ $T_{m}$ \textbf{and} $\delta_{\textrm{cost}}>\epsilon_{m}$
\textbf{do}

\quad{}$i$++;

\quad{}$\eta,\omega$ := \texttt{$\mathtt{addTrajectories}$}($\eta,\omega$);

\quad{}$\eta$ := \texttt{$\mathtt{deleteTrajectories}$}($\eta,\omega$);

\quad{}$\omega,\textrm{cost}$ := \texttt{$\mathtt{optimizePermutation}$}$(\eta,\omega)$;

\quad{}\textbf{if} $\textrm{cost}<\textrm{cost}^{*}$ \textbf{then} 

\quad{}\quad{}$\eta^{*}:=\eta$; $\omega^{*}$:= $\omega$; $\delta_{\textrm{cost}}:=$
$\textrm{cost}^{*}$- cost; $\textrm{ cost}^{*}$:= cost; 

Compute $\tau^{*}$ from $\eta^{*}$ and $\omega^{*}$ using (\ref{eq:mean_each_traj});

Transform $\hat{\boldsymbol{X}}:=\mathcal{R}(\eta^{*},\tau^{*})$;

\caption{Greedy algorithm for multi-object trajectory consensus with $T_{m}$
iterations and convergence threshold $\epsilon_{m}$.\protect\label{alg:bary-ospa}}
\end{algorithm}
\begin{algorithm}
\textbf{Input}: $\eta$, $\omega$.

\textbf{Output}: Updated $\omega$, cost.\vspace{-0.5em}

\rule[0.5ex]{1\columnwidth}{0.5pt}

Compute $\tau^{*}$ from $\eta$ and $\omega$ using (\ref{eq:mean_each_traj});

\textbf{for} $n:=1:N$ \textbf{do}

\quad{}$\omega(n,\cdot)$ := $\mathtt{OptimalAssignment}$$((\eta,\tau^{*}),(\xi^{(n)},\chi^{(n)}))$;

$\textrm{cost}:=S^{(r)}(\eta,\omega)$;

\caption{$\mathtt{optimizePermutation}$ function.\protect\label{alg:optim}}
\end{algorithm}

For each $n\in\{1:N\}$, the \texttt{$\mathtt{optimizePermutation}$}
function (Algorithm \ref{alg:optim}) is used to determine the optimal
existence assignment $\omega^{*}(n,\cdot)$. In this step, we first
determine the optimal trajectory list $\tau^{*}$ conditioned on $\eta$
and the current $\omega$ by solving (\ref{eq:mean_each_traj}) for
each $\ell\in\{1:L\}$. To obtain $\omega^{*}(n,\cdot)$, we use a
linear assignment solver \cite{jonker1987shortest} to assign elements
of the set $\{\ell:\eta_{\ell}=1\}$ to the set $\{\ell:\xi_{\ell}^{(n)}=1\}.$
The cost of assigning element $i$ to $j$ is $d_{\mathbb{T}}^{(c)}\!(\tau_{i}^{*},\chi_{j}^{(n)})$.
For any indices that remain unassigned or are not included in the
optimal assignments, we randomly assign them since any one of these
assignments yields the same cost.

The minimization on $\mathbb{B}^{L}$ is carried out by optimizing
each coordinate of $\eta$ through the functions \texttt{$\mathtt{addTrajectories}$}
(Algorithm \ref{alg:add-elem-ospa}) and\texttt{ $\mathtt{deleteTrajectories}$
}(Algorithm \ref{alg:del-elem-ospa}). The $\mathtt{addTrajectories}$
function updates each coordinate of $\eta$ from 0 to 1 whenever such
a change reduces the cost. The $\mathtt{deleteTrajectories}$ function
updates each coordinate of $\eta$ from 1 to 0 if doing so results
in a cost reduction. If the complexity of the algorithm to solve Problem
(\ref{eq:mean_each_traj}) is $\mathcal{O}(Z)$, the worst case complexity
of Algorithm \ref{alg:bary-ospa} is $\mathcal{O}(T_{m}L(NL^{2}+Z))$.
 In Figure \ref{fig:samples-mean-mo}, we show examples of multi-object
trajectory consensus of three point patterns computed by this algorithm.
\begin{algorithm}
\textbf{Input:} $\eta$, $\omega$.

\textbf{Output}: Updated $\eta$, $\omega$.\vspace{-0.5em}

\rule[0.5ex]{1\columnwidth}{0.5pt}

Compute $\tau^{*}$ from $\eta$ and $\omega$ using (\ref{eq:mean_each_traj});

\textbf{for} $n:=1:N$ \textbf{do}

\quad{} $\!\hat{\boldsymbol{U}}^{(n)}\!:=\!\{\chi_{\ell}^{(n)}|d_{\mathbb{T}}^{(c)}\!(\tau_{\ell}^{*},\chi_{\omega(n,\ell)}^{(n)})\!=\!c,$

\quad{} \quad{} \quad{} \quad{} \quad{} $\thinspace\thinspace\xi_{\omega(n,\ell)}^{(n)}=1,\eta_{\ell}=1,\ell=1:L\};$

\quad{}$\bar{\boldsymbol{U}}^{(n)}:=\!\{\chi_{\ell}^{(n)}|\xi_{\omega(n,\ell)}^{(n)}=1,\eta_{\ell}=0,\ell=1:L\}$;

\quad{}$\boldsymbol{U}^{(n)}:=\hat{\boldsymbol{U}}^{(n)}\cup\bar{\boldsymbol{U}}^{(n)}$;
$\boldsymbol{U}:=\cup_{n=1}^{N}\boldsymbol{U}^{(n)}$;

\quad{}$L_{A}:=\{\ell=1:L|\eta_{\ell}=0\}$;

\textbf{for} $\ell\in L_{A}$ \textbf{do}

\quad{}\textbf{if} $\boldsymbol{U}=\emptyset$ \textbf{then break}

\quad{}$u:=\mathtt{sampleOneElementUniformly}(\boldsymbol{U})$;
$\boldsymbol{U}:=\boldsymbol{U}-\{u\}$;

\quad{}$\acute{\omega}$ := $\mathtt{optimizeCluster}$($u$, $\omega$,
$\ell$, $\boldsymbol{U}^{(1:N)}$);

\quad{}$\acute{\eta}:=\eta$; $\acute{\eta}_{\ell}:=1;$

\quad{}\textbf{if} $S^{(r)}(\acute{\eta},\acute{\omega})<S^{(r)}(\eta,\omega)$
\textbf{then }$\eta:=\acute{\eta}$; $\omega:=\acute{\omega}$;

\rule[0.5ex]{1\columnwidth}{1pt}

$\mathtt{optimizeCluster}$ sub-function

\textbf{Input}: $u$, $\omega$, $\ell$, $\boldsymbol{U}^{(1:N)}$.

\textbf{Output}: $\acute{\omega}$.\vspace{-0.5em}

\rule[0.5ex]{1\columnwidth}{0.5pt}

$\acute{\omega}:=\omega$;

\textbf{for} $n:=1:N$ \textbf{do}

\quad{}$j:=\arg\min{}_{\acute{j}\in\{1:|\boldsymbol{U}^{(n)}|\}}d_{\mathbb{T}}^{(c)}(u,\boldsymbol{U}_{\acute{j}}^{(n)})$;

\quad{}$i:=\mathtt{originalIndex}(j)$; (index of $\boldsymbol{U}_{j}^{(n)}$
in $\chi^{(n)}$)

\quad{}$\textrm{Find }\acute{\ell}\textrm{ such that }\acute{\omega}(n,\acute{\ell})=i$;

\quad{}$\acute{i}:=\acute{\omega}(n,\ell)$; $\acute{\omega}(n,\ell):=i$;\ $\acute{\omega}(n,\acute{\ell}):=\acute{i}$;

\caption{\texttt{$\mathtt{addTrajectories}$} function.\protect\label{alg:add-elem-ospa}}
\end{algorithm}
\begin{algorithm}
\textbf{Input}: $\eta$, $\omega$

\textbf{Output}: Updated $\eta$.\vspace{-0.5em}

\rule[0.5ex]{1\columnwidth}{0.5pt}

\textbf{for} $\ell:=1:L$ \textbf{do}

\quad{}\textbf{if} $\eta_{\ell}=1$ \textbf{do }$\acute{\eta}:=\eta$;
$\acute{\eta}_{\ell}:=0;$

\quad{}\quad{}\textbf{if} $S^{(r)}(\acute{\eta},\omega)<S^{(r)}(\eta,\omega)$
\textbf{then }$\eta:=\acute{\eta}$;

\caption{\texttt{$\mathtt{deleteTrajectories}$} function.\protect\label{alg:del-elem-ospa}}
\end{algorithm}

\section{Fr\'{e}chet Mean via Gibbs Sampling \protect\label{sec:Gibbs-sampling}}

While the greedy algorithms proposed in the previous subsections are
inexpensive and well-suited for online applications, convergence to
the global optima is an open question. In the absence of theoretical
results, we endeavor to offer some empirical analysis by benchmarking
the greedy solutions against near-optimal solutions (in the context
of trajectory and multi-object trajectory consensus). To this end,
we propose a stochastic optimization technique that can reach optimality,
albeit at the expense of increased computations. 

Our proposed optimization method uses Gibbs sampling to draw samples
from a distribution whose peaks align with the minima of the cost
function, and retains samples with the least costs. Specifically,
given a cost function $g:\mathbb{Y}\rightarrow[0,\infty)$ and a constraint
function $F:\mathbb{Y}\rightarrow\{0,1\}$, we define a distribution
\begin{equation}
\rho(y)\propto e^{-\alpha(g(y))}F(y),\label{eq:gibbs-sample-dist-1}
\end{equation}
on $\mathbb{Y}$, where $\alpha$ is some positive constant. Since
$e^{-\alpha(\cdot)}$ is bounded, and monotonically decreasing on
$[0,\infty)$, the maxima of $\rho$ coincide with the minima of $g$
subject to the constraint $F(y)=1$ (this also holds if we replace
$e^{-\alpha(\cdot)}$ by any bounded monotonically decreasing function).
Hence, all samples from $\rho$ satisfy the constraint $F(y)=1$ and
the most likely samples are the minima of $g$. 

The Gibbs sampler generates a Markov chain, wherein a new iterate
$\acute{y}=(\acute{y}_{1:K})$ is generated from the current iterate
$y=(y_{1:K})$ by sampling in each coordinate from the corresponding
conditional distribution $\rho_{k}(\acute{y}_{k}|\acute{y}_{1:k-1},y_{k+1:K})$
as follows \cite{casella1992explaining}
\begin{align*}
\acute{y}_{1} & \sim\rho_{1}(\cdot|y_{2:K})\\
\acute{y}_{2} & \sim\rho_{2}(\cdot|\acute{y}_{1},y_{3:K})\\
 & \thinspace\thinspace\thinspace{\scriptstyle \vdots}\\
\acute{y}_{K} & \sim\rho_{K}(\cdot|\acute{y}_{1:K-1}).
\end{align*}
Choosing the conditional distributions
\begin{equation}
\rho_{k}(\acute{y}_{k}|\acute{y}_{1:k-1},y_{k+1:K})\propto\rho(\acute{y}_{1:k},y_{k+1:K}),\label{eq:gibbs-sample-dist-1-1}
\end{equation}
and assuming the chain can visit all possible states in a finite number
of steps, it can be shown that the Markov chain converges to the stationary
distribution $\rho$, i.e., when the chain runs sufficiently long,
the subsequent iterates are distributed according to $\rho$. Moreover,
starting from an initial $y\in\mathbb{Y}$, the Gibbs sampler converges
to the distribution $\rho$ at an exponential rate \cite{gallager2013stochastic}.
In Subsections \ref{sec:Gibbs-Sampling-Trajectory} and \ref{sec:Gibbs-mo-trajectory},
we adapt the Gibbs sampler, respectively, to compute the mean trajectory
and mean multi-object trajectory.

\subsection{Gibbs Sampling for Mean Trajectory\protect\label{sec:Gibbs-Sampling-Trajectory}}

In trajectory consensus, $\mathbb{Y}=\mathbb{B}^{L}$, $y=\gamma$,
the cost function $g(y)=W^{(r)}(\gamma)$, and there is no constraint.
For completeness, the pseudocode to compute the Fr\'{e}chet mean trajectory
is given in Algorithm \ref{alg:gibbs-existence}, and convergence
of the Gibbs sampler (to the stationary distribution at an exponential
rate) directly follows from Lemma 4.3.4 of \cite{gallager2013stochastic}.
If the complexity of the algorithm to solve Problem (\ref{eq:xk-star})
is $\mathcal{O}(M)$, the complexity of Algorithm \ref{alg:gibbs-existence}
is $\mathcal{O}(G_{t}K^{2}M)$. 
\begin{algorithm}
\textbf{Output}: Mean trajectory $\hat{v}$.\vspace{-0.5em}

\rule[0.5ex]{1\columnwidth}{1pt}

Initialize $\gamma:=\boldsymbol{0}$;

Set $\gamma_{k}:=1$ if $\sum_{n=1}^{N}\mathbf{1}_{\mathcal{D}_{\!v^{(n)}}}^{(k)}\!>\!0$;

$\gamma^{*}:=\gamma$; $\textrm{\ensuremath{\text{cost}^{*}}}:=$
$W^{(r)}(\gamma^{*})$;

\textbf{for} $t:=1:G_{t}$ \textbf{do}

\quad{}\textbf{for} $k:=1:K$ \textbf{do}

\quad{}\quad{}$\gamma_{k}\sim\mathtt{{\small \mathtt{CatSampling}}}(\{0,1\},\rho_{k}(\cdot|\gamma_{1:k-1},\gamma_{k+1:K}))$;

\quad{}\quad{}\textbf{if} $W^{(r)}(\gamma)<\textrm{\ensuremath{\text{cost}^{*}}}$
\textbf{then} $\gamma^{*}:=\gamma$; $\textrm{\ensuremath{\text{cost}^{*}}}:=W^{(r)}(\gamma)$;

Compute $x^{*}$ from $\gamma^{*}$ using (\ref{eq:xk-star});

Transform $\hat{v}:=\mathcal{R}(\gamma^{*},x^{*})$;

\caption{Gibbs sampling for trajectory consensus with $G_{t}$ iterations and
categorical sampler (denoted as $\mathtt{{\small CatSampling}}$).\negthinspace{}\negthinspace{}\negthinspace{}
\protect\label{alg:gibbs-existence}}
\end{algorithm}

Similar to the greedy algorithm introduced in Section \ref{subsec:Greedy-Algorithm-Trajectory},
we can approximate the cost $W^{(r)}(\cdot)$ by using $x^{*}$ computed
by (\ref{eq:approx-x}) in place of $x^{*}(\gamma)$. This approximation
reduces the complexity of Algorithm \ref{alg:gibbs-existence} to
$\mathcal{O}(K(G_{t}+M))$. 

\subsection{Gibbs Sampling for Mean Multi-Object Trajectory\protect\label{sec:Gibbs-mo-trajectory}}

In multi-object trajectory consensus, $\mathbb{Y}=\mathbb{B}^{L}\times\left[\Pi_{L}\right]^{N}$,
$y_{1:L}=\eta_{1:L}$, $y_{L+1:K}=\omega^{(1:N)}$, $K=L+N$, the
cost function $g(y_{1:K})=S^{(r)}(\eta_{1:L},\omega^{(1:N)})$ and
the constraint function $F(y_{1:K})=\prod_{n=1}^{N}\boldsymbol{1}_{\Omega^{(n)}(\eta_{1:L})}(\omega^{(n)})$,
because each $\omega^{(n)}$ is an existence assignment, i.e., $\omega^{(n)}\in\Omega^{(n)}(\eta_{1:L})$.
Hence, the stationary distribution is 
\[
\rho(\eta_{1:L},\omega^{(1:N)})\propto e^{-\alpha(S^{(r)}(\eta_{1:L},\omega^{(1:N)}))}\prod_{n=1}^{N}\boldsymbol{1}_{\Omega^{(n)}(\eta_{1:L})}(\omega^{(n)}).
\]
Since each $\omega^{(n)}$ is an existence assignment, i.e., a permutation
vector in $\Pi_{L}$ with certain properties, sampling $\acute{\omega}^{(n)}$
from the corresponding conditional
\begin{equation}
\rho_{L+n}(\cdot|\acute{\eta}_{1:L},\acute{\omega}^{(1:n-1)},\omega^{(n+1:N)}),\label{eq:assignment_cond}
\end{equation}
is more involved than sampling the existence variable $\eta_{\ell}\in\mathbb{B}=\{0,1\}$.
Sampling $\acute{\omega}^{(n)}$ coordinate by coordinate is difficult
because each component of the permutation vector must be different
from the others. Using block Gibbs sampling is expensive when the
block size is large. To circumvent this issue, we propose a reparameterization
technique based on \textit{repeated insertion} \cite{doignon2004repeated}
that enables efficient Gibbs sampling. 

Given a \textit{reference} permutation vector $\vartheta=[\vartheta_{1},...,\vartheta_{L}]$,
an\textit{ insertion vector} is a vector of positive integers $j=[j_{1},...,j_{L}]$
that satisfies $j_{\ell}\leq\ell$, $\forall\ell\leq L$. A \textit{repeated
insertion function} $\varrho_{\vartheta}$ maps an insertion vector
$j$ to a permutation vector $\pi$, i.e., $\pi=\varrho_{\vartheta}(j)$,
by inserting $\vartheta_{\ell}$ into position $j_{\ell}$ of $\pi$
and shifting its previous value (if there is one) to position $j_{\ell}+1$,
for each $\ell\leq L$. This is illustrated below.

\begin{example}Consider a reference permutation vector $\vartheta=[3,1,2,4]$
and an insertion vector $j=[1,2,2,3]$. We start with an empty vector
$\pi=[\Circle,\Circle,\Circle,\Circle]$. We first insert $\vartheta_{1}=3$
to position $j_{1}=1$ of $\pi$ which yields $\pi=[3,\Circle,\Circle,\Circle]$.
We then insert $\vartheta_{2}=1$ to position $j_{2}=2$ of $\pi$
which yields $\pi=[3,1,\Circle,\Circle]$. We continue to insert $\vartheta_{3}=2$
to position $j_{3}=2$ of $\pi$. But since position 2 of $\pi$ is
currently occupied by 1, we shift 1 to position 3 and insert $\vartheta_{3}=2$
into position 2, which yields $\pi=[3,2,1,\Circle]$. Finally we insert
$\vartheta_{4}=4$ to position $j_{4}=3$. But since position 3 is
currently occupied by 1, we shift 1 to position 4 and insert 4 in
position 3, which finally yields $\pi=[3,2,4,1]$.\end{example} 

Noting that the repeated insertion function $\varrho_{\vartheta}$
is bijective \cite{lu2014effective}, we can transform the problem
of sampling a permutation vector $\pi$ from a distribution $p(\cdot)$
(such as the conditional (\ref{eq:gibbs-sample-dist-1-1})), into
sampling an insertion vector $j$ from $\tilde{p}_{\vartheta}(\cdot)\triangleq p(\varrho_{\vartheta}(\cdot))$
and then recover $\pi=\varrho_{\vartheta}(j)$. Sampling insertion
vectors from $\tilde{p}_{\vartheta}$ can be accomplished coordinate
by coordinate via Gibbs sampling with conditionals
\[
\tilde{p}_{\vartheta,\ell}(\acute{j}_{\ell}|\acute{j}_{1:\ell-1},j_{\ell+1:L})\propto\tilde{p}_{\vartheta}(\acute{j}_{1:\ell-1},\acute{j}_{\ell},j_{\ell+1:L}).
\]
Specifically, for sampling $\omega^{(n)}$ from (\ref{eq:assignment_cond}),
the above conditional is proportional to
\[
\rho(\acute{\eta}_{1:L},\acute{\omega}^{(1:n-1)},\varrho_{\vartheta}([\acute{j}_{1:\ell},j_{\ell+1:L}]),\omega^{(n+1:N)}).
\]

The pseudocode to compute the mean multi-object trajectory via Gibbs
sampling is provided in Algorithm \ref{alg:bary-ospa-samp}, and
convergence of the Gibbs sampler (to the stationary distribution at
an exponential rate) is proven in SM Subsection \ref{subsec:Convergence-Property}.
If the complexity of the algorithm to solve Problem (\ref{eq:mean_each_traj})
is $\mathcal{O}(Z)$, the complexity of Algorithm \ref{alg:bary-ospa-samp}
is $\mathcal{O}(G_{m}NL^{3}Z)$, where $G_{m}$ is the number of
iterates of the Gibbs sampler. Implementation strategies to speed
up computations are discussed in SM Subsection \ref{subsec:Enhancement-Strategies}.
\begin{remark}Sampling distributions on the space of permutations
is an important problem in computational statistics \cite{pitman2019regenerative}.
The proposed algorithm is the first to exploit repeated insertion
to perform Gibbs sampling on the space of permutations.\end{remark}
\begin{algorithm}
\textbf{Output}: Mean multi-object trajectory $\hat{\boldsymbol{X}}$.\vspace{-0.5em}

\rule[0.5ex]{1\columnwidth}{0.5pt}

Initialize $\eta,\omega$ using Algorithm \ref{alg:bary-ospa};

$\eta^{*}:=\eta$; $\omega^{*}$ := $\omega$; $\text{cost}^{*}$:=
$S^{(r)}(\eta,\omega)$;

\textbf{for} $g:=1:G_{m}$ \textbf{do}

\quad{}{\small\texttt{/{*}sample existence list{*}/}}{\small\par}

\quad{}\textbf{for} $\ell:=1:L$ \textbf{do}

\quad{}\quad{}$\eta_{\ell}\sim{\small \mathtt{CatSampling}}(\{0,1\},\rho_{\ell}(\cdot|\eta_{1:\ell-1},\eta_{\ell+1:L},\omega))$;

\quad{}\quad{}\textbf{if} $S^{(r)}(\eta,\omega)<\text{cost}^{*}$\textbf{
then}

\quad{}\quad{}\quad{}$\eta^{*}:=\eta$; $\omega^{*}$:= $\omega$;
$\text{cost}^{*}:=S^{(r)}(\eta,\omega)$;

\quad{}{\small\texttt{/{*}sample existence assignments{*}/}}{\small\par}

\quad{}\textbf{for} $n:=1:N$ \textbf{do}

\quad{}\quad{}$j:=\varrho_{\vartheta}^{-1}(\omega^{(n)})$;

\quad{}\quad{}\textbf{for} $\ell:=1:L$ \textbf{do}

\quad{}\quad{}\quad{}$j_{\ell}\sim\mathtt{{\small \mathtt{CatSampling}}}(\{1:\ell\},\tilde{p}_{\vartheta,\ell}(\cdot|j_{1:\ell-1},j_{\ell+1:L}))$;

\quad{}\quad{}\quad{}$\omega:=[\omega^{(1)},...,\omega^{(n-1)},\varrho_{\vartheta}(j),\omega^{(n+1)},...,\omega^{(N)}]$;

\quad{}\quad{}\quad{}\textbf{if} $S^{(r)}(\eta,\omega)<\text{cost}^{*}$
\textbf{then}

\quad{}\quad{}\quad{}\quad{}$\eta^{*}:=\eta$; $\omega^{*}$:=
$\omega$; $\text{cost}^{*}:=S^{(r)}(\eta,\omega)$;

Compute $\tau^{*}$ from $\eta^{*}$ and $\omega^{*}$ using (\ref{eq:mean_each_traj});

Transform $\hat{\boldsymbol{X}}:=\mathcal{R}(\eta^{*},\tau^{*})$;

\caption{Gibbs sampling for multi-object trajectory consensus with $G_{m}$
iterations.\protect\label{alg:bary-ospa-samp}}
\end{algorithm}

\section{Experimental Results\protect\label{sec:Experimental-Results}}

This section presents the first practical demonstration of Fr\'{e}chet
mean (FM) multi-object trajectory consensus, that is, the tractable
computation of the geometric barycenter or ``average'' for multiple
sets of trajectory valued elements. The results explore the behavior
of FM consensus with the OSPA\textsuperscript{(2)} metric (defined
in (\ref{eq:ospa^(2)})), where $p=100$m and $c=100\sqrt{2}$m. We
demonstrate the results for the first-order ($r=1$) and the second-order
($r=2$) Fr\'{e}chet means, denoted as FM\textsuperscript{(1)} and FM\textsuperscript{(2)}
respectively. Both of the proposed implementations are considered,
namely greedy search as in Sections \ref{sec:track-dist}-\ref{sec:mt-consensus}
(with 5 and 100 iterations respectively for trajectory and multi-object
trajectory consensus) as well as Gibbs sampling as in Section \ref{sec:mt-consensus}
(with 100 and 10 iterations respectively for trajectory and multi-object
trajectory consensus). 
\begin{figure}
\begin{centering}
\includegraphics[bb=15bp 0bp 930bp 709bp,width=1\columnwidth]{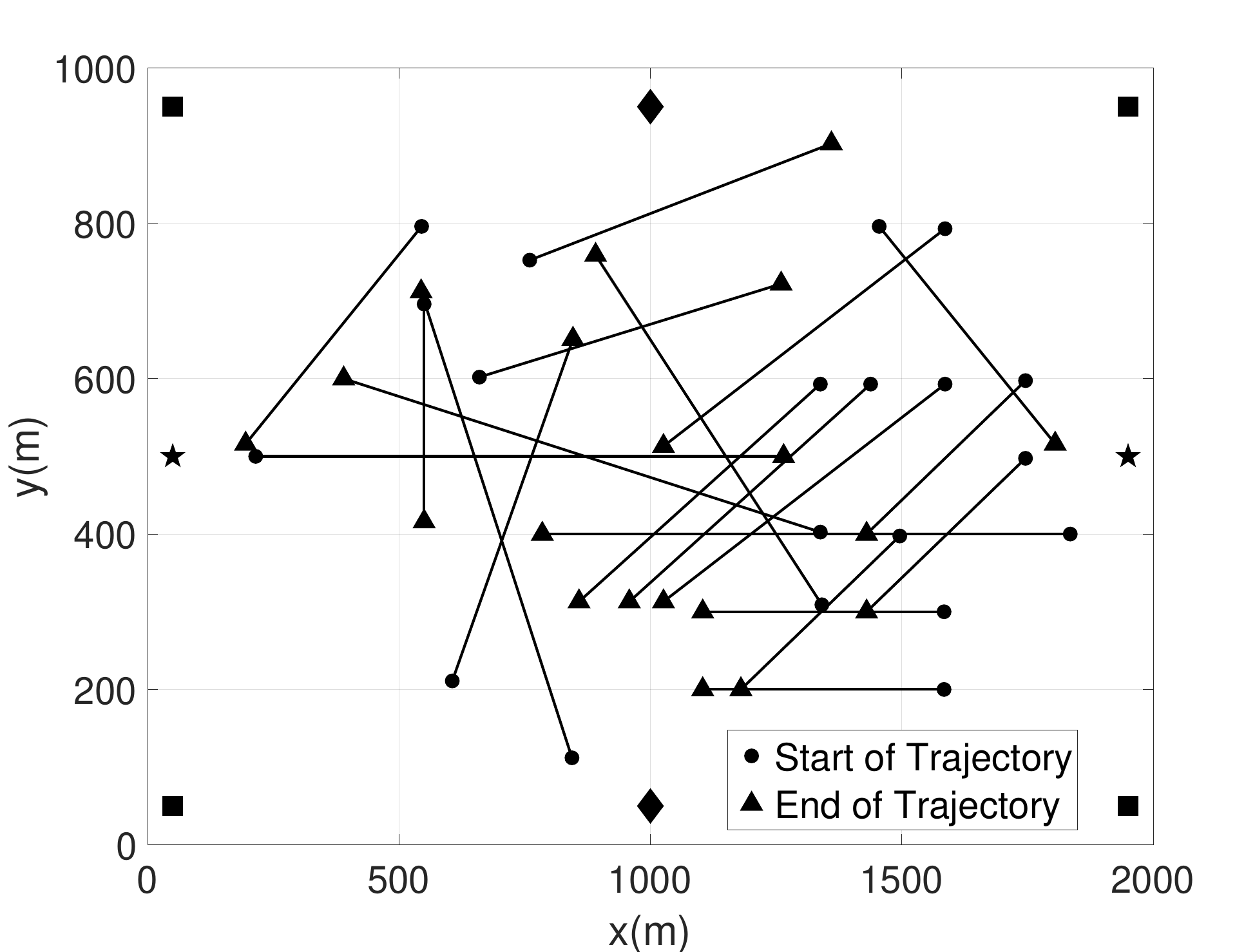}
\par\end{centering}
\caption{True trajectories and network configuration where the nodes are annotated
by \textifsymbol[ifgeo]{96}, \FiveStar{} or {\large\ding{169},} see
Table \ref{tab:sensor-nodes} for details of the sensor nodes.\protect\label{fig:scenario}}
\end{figure}
\begin{table}
\caption{Summary of sensor types. The parameters $P_{D}$ and $\lambda_{c}$
are respectively the sensor detection probability and clutter rate.
\protect\label{tab:sensor-nodes}}

\centering{}%
\begin{tabular}{cccccc}
\toprule 
{\small Type} & {\small Symbol} & {\small Sensor} & {\small Tracker} & {\small$P_{D}$} & {\small$\lambda_{c}$}\tabularnewline
\midrule
\midrule 
{\small 1} & {\small\textifsymbol[ifgeo]{96}} & {\small R-RR-B} & {\small TO-MHT} & {\small 0.85} & {\small 50}\tabularnewline
\midrule 
{\small 2} & {\small\ding{169}} & {\small POS} & {\small LMB} & {\small 0.7} & {\small 10}\tabularnewline
\midrule 
{\small 3} & {\small\FiveStar{}} & {\small B-R} & {\small GLMB} & {\small 0.9} & {\small 70}\tabularnewline
\bottomrule
\end{tabular}
\end{table}

Direct comparisons are also undertaken with the two latest approaches
for trajectory consensus, DBSCAN clustering for tracks (DBSCAN-T)
\cite{he2018multi} and Track Consensus (TC) in \cite{nguyen2021distributed}.
It should be noted that the proposed FM consensus is inherently a
multi-scan or window-based formulation (batch) whereas DBSCAN-T and
TC are both single-scan or recursive formulations (scan-by-scan).
Thus, the FM approach also marks the first tractable implementation
of a batch as opposed to scan-by-scan algorithm for multi-object trajectory
consensus. 
\begin{figure*}
\begin{centering}
\includegraphics[bb=0bp 0bp 900bp 709bp,clip,width=0.357\textwidth]{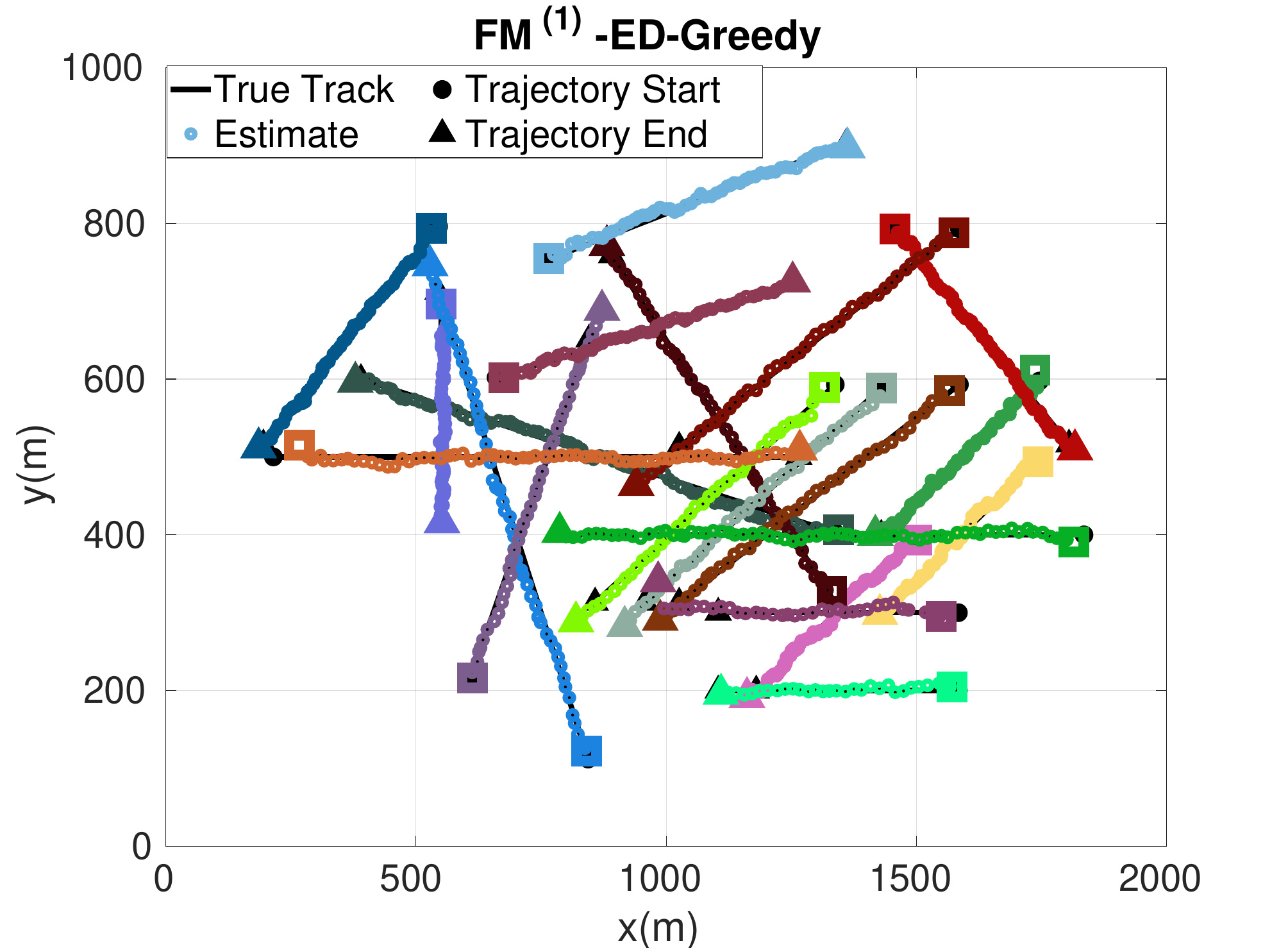}\includegraphics[bb=70bp 0bp 880bp 709bp,clip,width=0.321\textwidth]{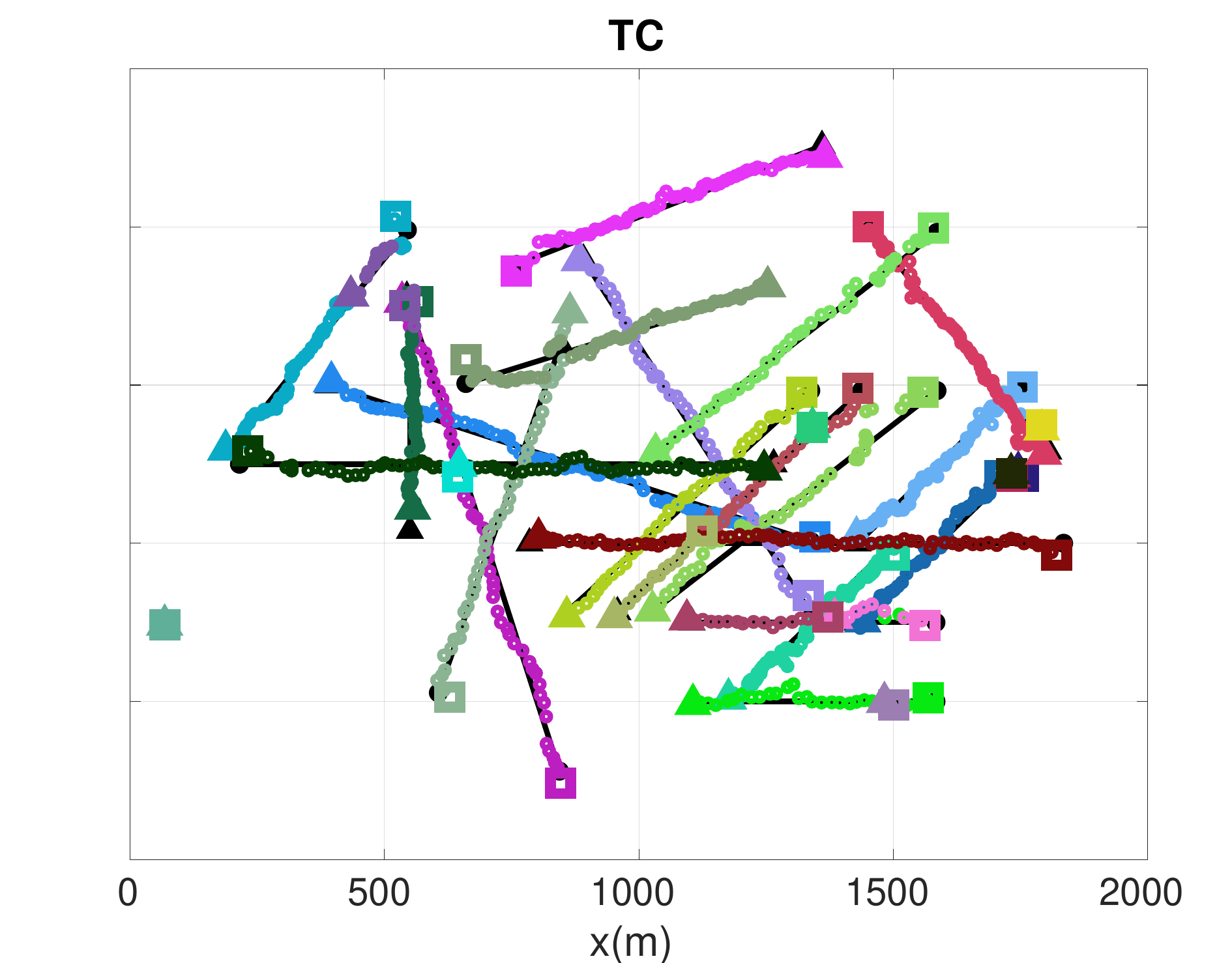}\includegraphics[bb=70bp 0bp 880bp 709bp,clip,width=0.321\textwidth]{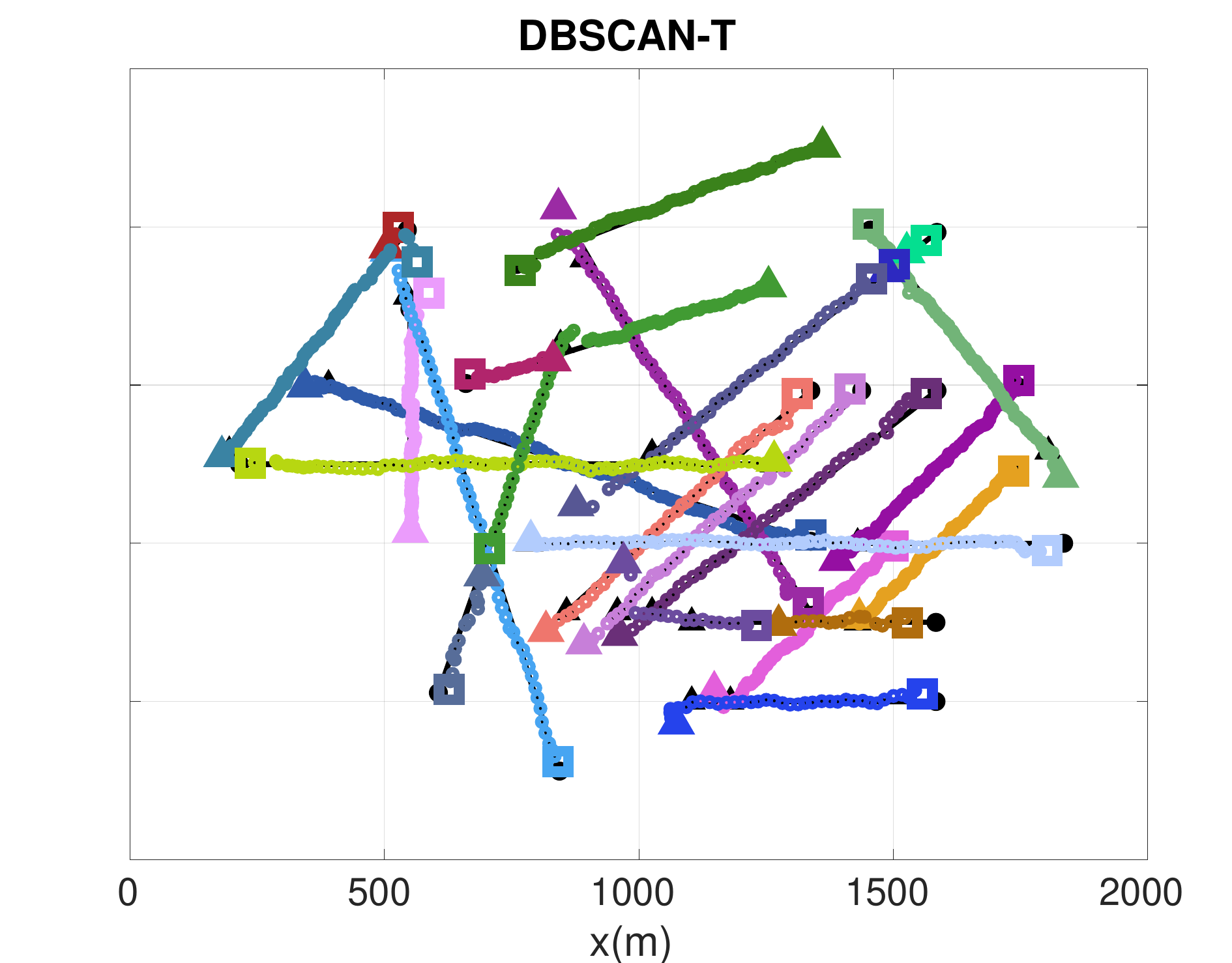}
\par\end{centering}
\caption{Sample outputs from FM\protect\textsuperscript{(1)} with Euclidean
distance and greedy search, TC and DBSCAN-T.\protect\label{fig:case1-est}}
\end{figure*}

A networked multi-object tracking scenario is employed to generate
a synthetic dataset for the purpose of demonstrating multi-object
trajectory consensus. A fixed ground truth with multiple sensor nodes
is used to generate multiple sets of imperfect estimates. In general,
any mix of sensor types or even human mediated reports can be processed
since the FM or consensus is not dependent on the sensor types or
models. The sets of node outputs are collectively fed into the consensus
algorithm whose result is then compared to the ground truth. 

The ground truth has 20 trajectories of different lengths over 100
time scans of 1s intervals. The individual node reports come from
8 distinctly located sensors, each chosen from 3 different types,
and producing estimates of the multi-object trajectory ground truth
with varying accuracy. The ground truths and sensor locations are
shown in Figure \ref{fig:scenario}. The multi-object trajectory estimate
from each node is simulated as the output of various tracking filters
run on different measurement types. In particular, the measurement
types are range, range-rate and bearing (R-RR-B), bearing range (B-R)
and 2D position (POS) with noise standard deviation of 20m on position
and range, $1^{\circ}$ on bearing noise, and 2m/s on range rate.
The trackers are track-oriented MHT (TO-MHT) \cite{blackman1999design},
LMB \cite{reuter2014labeled}, or GLMB filters \cite{vo2013labeled}.
All trackers use a constant velocity motion model with a sampling
time of 1s and process noise standard deviation of 5m/s\textsuperscript{2}.
Node information is summarized in Table \ref{tab:sensor-nodes}. 
\begin{figure}
\begin{centering}
\includegraphics[width=1\columnwidth]{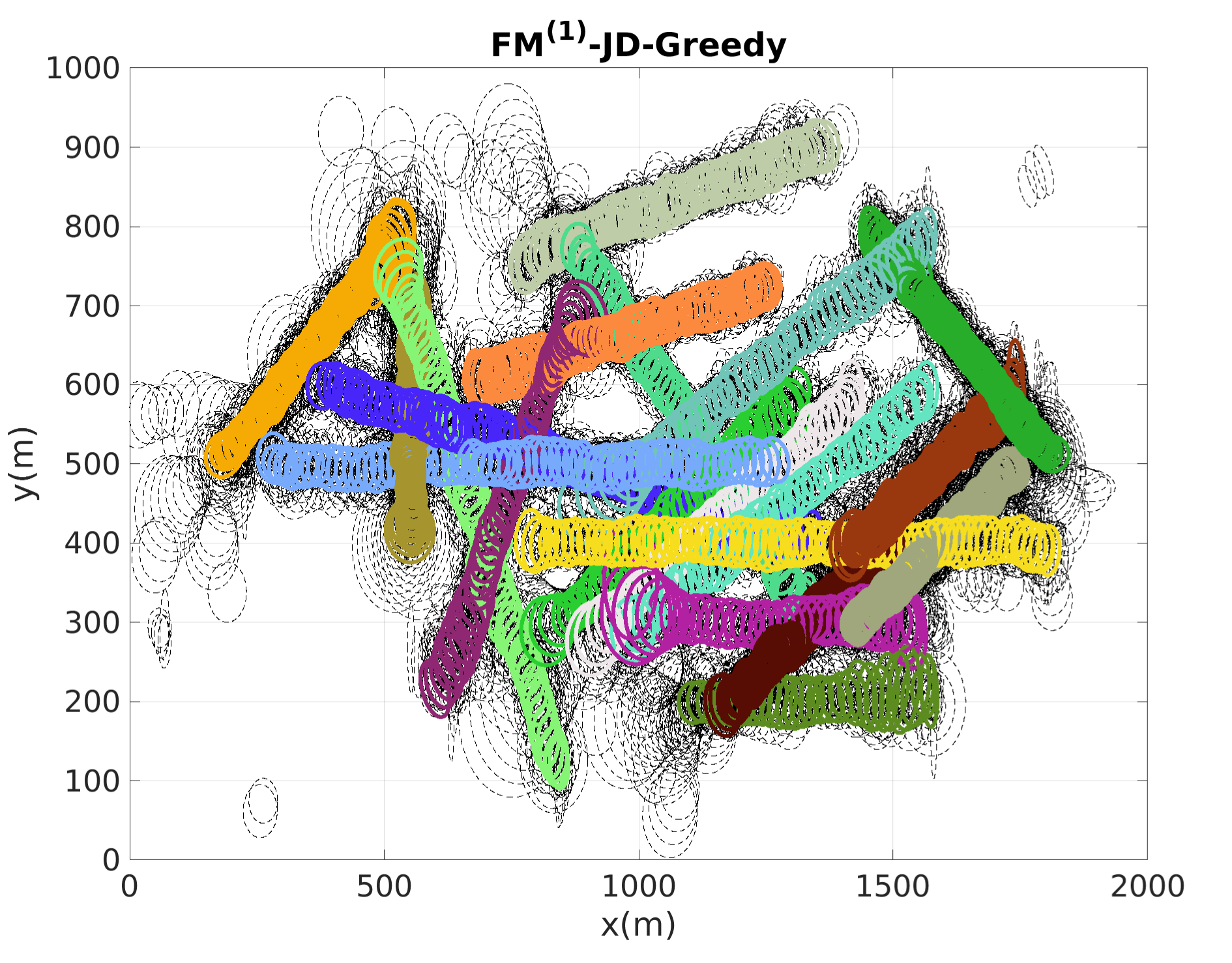}
\par\end{centering}
\caption{Sample output from FM\protect\textsuperscript{(1)} fusion with uncertainty.
The ellipses represent 95\% confidence intervals of the Gaussian distributions.
Dashed line ellipses are estimates from the nodes and solid ellipses
are the fused results. \protect\label{fig:case1-jd-fuse}}
\end{figure}

In addition to deterministic estimates (wherein the Euclidean distance
(ED) is used as the base distance), we also apply our technique to
fuse multi-object trajectories with uncertainty. Presumably, the reported
trajectories from each node are represented by Gaussian distributions,
i.e., if a trajectory exists at time $k$, its state is a Gaussian
distribution parameterized by a mean vector and a covariance matrix.
To illustrate the application of our technique in Fr\'{e}chet-mean fusion
of tracks with uncertainty, we use the techniques proposed in \cite{nielsen2009sided}
to compute the centroids of Gaussians w.r.t. Jeffrey's divergence
(JD)\footnote{JD, which is a symmetrized KLD, is a pseudo-metric since it does not
satisfy the triangle inequality. We use JD in our demonstration owing
to the popularity of KLD for fusing distributions via the geometric
mean.}. Since computing the centroid Gaussian distributions is expensive,
we only compute the Fr\'{e}chet mean multi-object trajectory (with uncertainty)
using the greedy search method.

A sample output from each of FM\textsuperscript{(1)}\negthinspace{}
(Greedy) via ED, TC and DBSCAN-T is given in Figure \ref{fig:case1-est}.
Further, we also illustrate the output from FM\textsuperscript{(1)}\negthinspace{}
fusion with uncertainty via JD in Figure \ref{fig:case1-jd-fuse}.
It can be seen that outputs of the FM approaches correctly identify
all trajectories. In contrast, the TC output shows some susceptibility
to track switching and fragmentation, while DBSCAN-T additionally
exhibits late initiation and termination. These observations are consistent
with the fact that FM\textsuperscript{(1)}\negthinspace{} is a multi-scan
or window-based formulation which processes batch data, whereas TC
and DBSCAN-T are single-scan or recursive strategies which cannot
change past estimates. 

\begin{figure}
\begin{centering}
\includegraphics[width=0.95\columnwidth]{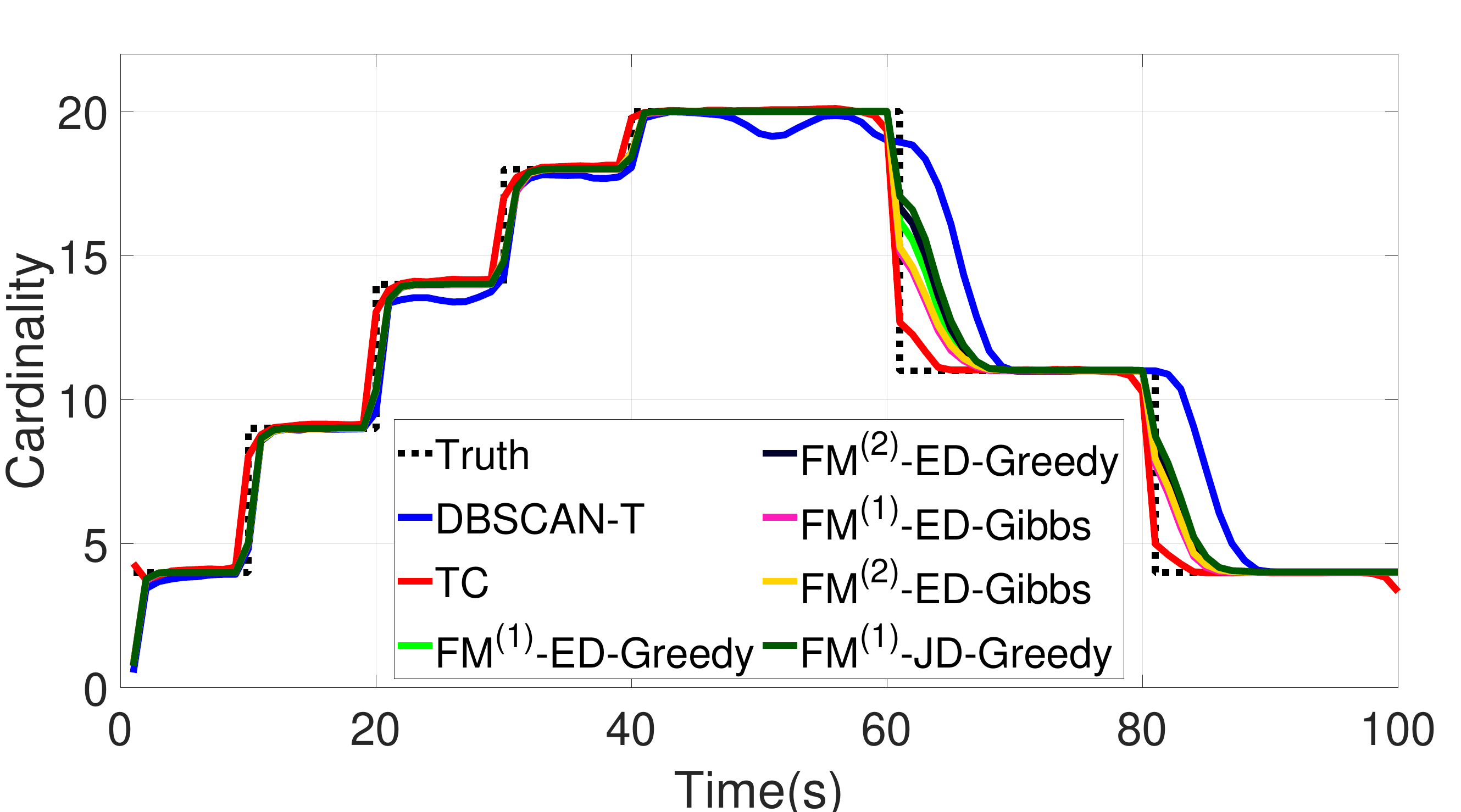}
\par\end{centering}
\caption{Average estimated cardinality. \protect\label{fig:case1-avg-card}}
\end{figure}

\begin{figure}
\begin{centering}
\includegraphics[width=1\columnwidth]{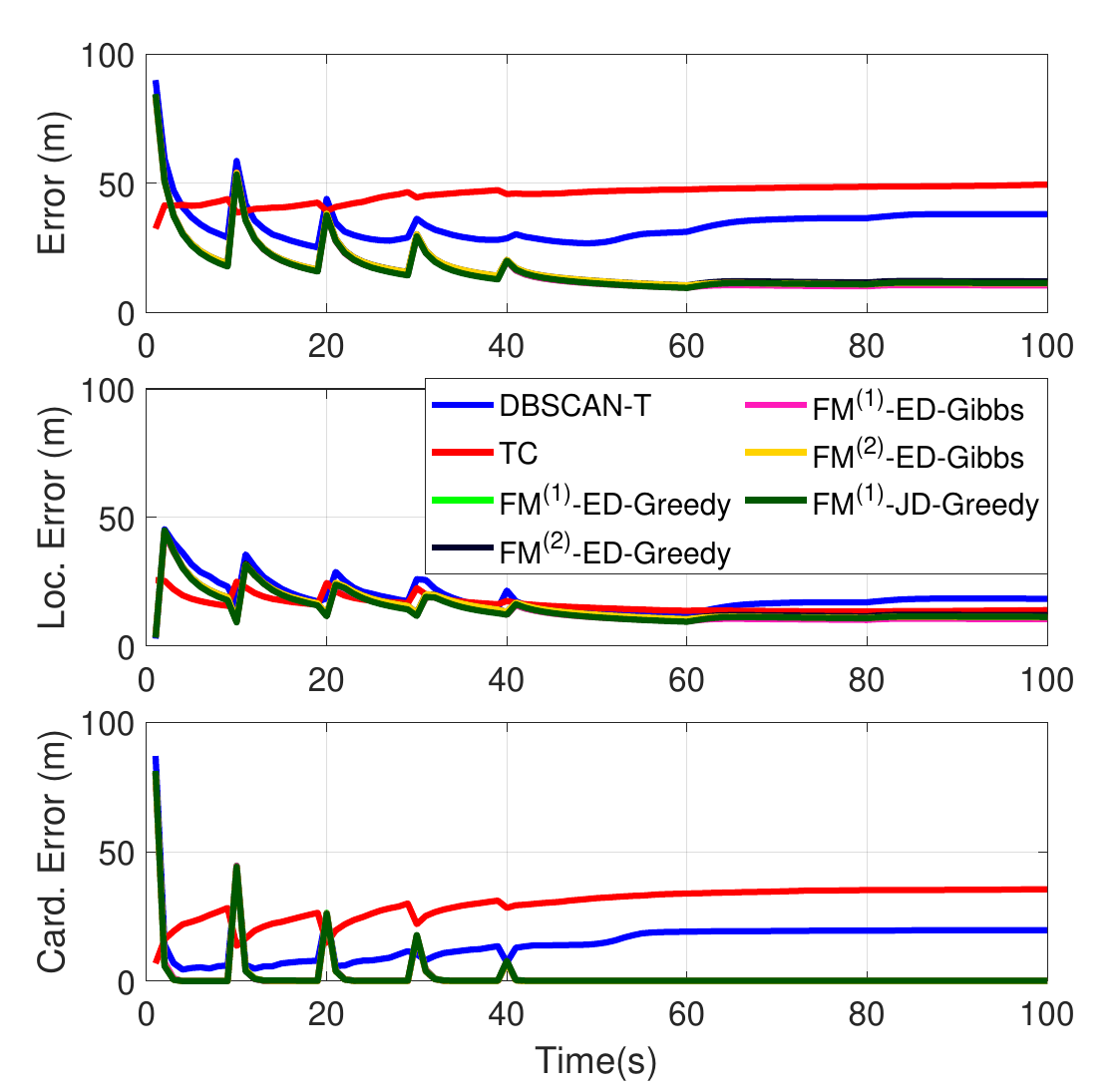}
\par\end{centering}
\caption{Average OSPA\protect\textsuperscript{(2)} over all MC runs at different
time steps. All FM-based solutions have almost identical performance.\protect\label{fig:case1-avg-ospa2-gw}}
\end{figure}

To confirm these observations, 100 Monte Carlo trials are performed.
The instantaneous estimated cardinality versus time is shown in Figure
\ref{fig:case1-avg-card}. The OSPA\textsuperscript{(2)} metric \cite{beard2020solution},
with order of 1, cut-off distance and cardinality penalty of 100m,
is calculated on an expanding window and shown versus time in Figure
\ref{fig:case1-avg-ospa2-gw}. Importantly, the OSPA\textsuperscript{(2)}
is a formal distance function which jointly penalizes localization
and cardinality errors, in addition to trajectory fragmentation and
labeling errors. The OSPA\textsuperscript{(2)} error value at a given
instant can be interpreted as a time-averaged per-trajectory error,
which in this study is calculated over an expanding window from the
scenario start to the current time. 

The cardinality curves in Figure \ref{fig:case1-avg-card} indicate
that FM consensus generally results in correct cardinality estimates
with quick track initiation but some delay in track termination. This
behavior reflects the output of the underlying tracking filters at
each of the sensor nodes since they all exhibit the same characteristics
which are reported as the consensus. The OSPA\textsuperscript{(2)}
curves in Figure \ref{fig:case1-avg-ospa2-gw} further indicate that
the greedy search (for both FM\textsuperscript{(1)} and FM\textsuperscript{(2)}),
using either ED and JD as base distances, and Gibbs sampling implementations
have almost identical performance. Importantly, the FM results in
the lowest OSPA\textsuperscript{(2)} error of the three approaches
since it implements window-based or multi-scan consensus scheme. 

While TC has very accurate instantaneous cardinality estimates, the
peer-to-peer nature exacerbates errors due to track switching and
fragmentation, as indicated by the higher OSPA\textsuperscript{(2)}
error. A similar examination of DBSCAN-T confirms the noticeable lag
in track initialization and termination which occurs in addition to
track switching and fragmentation. Even though DBSCAN-T is a centralized
scheme, the single scan recursive nature combined with its generation
of singleton clusters, results in more overall errors. 
 
Further analysis is presented in Table \ref{tab:case1-avg-error}
which shows the final OSPA\textsuperscript{(2)} error calculated
over the entire scenario along with the total execution time. The
final OSPA\textsuperscript{(2)} errors for each of the node estimates
are also shown in Table \ref{tab:case1-ospa2-per-node}. The final
OSPA\textsuperscript{(2)} values confirm that FM consensus results
in the lowest estimation error and significantly improves the raw
sensor node outputs. Further, the greedy search implementation is
marginally less accurate than Gibbs sampling but is orders of magnitude
faster. Among the FM consensus algorithms, the second-order FM techniques
exhibit slightly worse performance than their first-order counterparts,
while demonstrating a noticeable reduction in run-time. The better
performance of the order-1 FM could be attributed to the resilience
of the median against noise. The more efficient order-2 FM (with the
Euclidean base distance) is due to the simpler computation of the
fused trajectory state at each time step by averaging the corresponding
data points, whereas computing the median requires an iterative process.
Overall, the results suggest that the FM via greedy search produces
significant improvements in consensus performance compared to existing
approaches and with relatively efficient computation times.
\begin{table}
\caption{Average error over all MC runs. One standard deviation shown in parenthesis.\protect\label{tab:case1-avg-error}}

\centering{}{\small{}%
\begin{tabular}{lll}
\toprule 
{\small\textbf{Method}} & {\small\textbf{OSPA\textsuperscript{\textbf{(2)}}(m)}} & {\small\textbf{Exec. Time(s)}}\tabularnewline
\midrule
\midrule 
{\small DBSCAN-T} & {\small 38.05(5.38)} & {\small 39.19(0.93)}\tabularnewline
{\small TC} & {\small 49.47(8.50)} & {\small 397.36(7.24)}\tabularnewline
{\small FM\textsuperscript{(1)}-ED (Greedy)} & {\small 11.05(0.80)} & {\small 27.57(2.29)}\tabularnewline
{\small FM\textsuperscript{(2)}-ED (Greedy)} & {\small 12.05(1.05)} & {\small\textbf{\uline{21.55(1.66)}}}\tabularnewline
{\small FM\textsuperscript{(1)}-ED (Gibbs)} & {\small\textbf{\uline{10.48(0.81)}}} & {\small 1099.52(76.84)}\tabularnewline
{\small FM\textsuperscript{(2)}-ED (Gibbs)} & {\small 11.21(0.81)} & {\small 1073.81(61.44)}\tabularnewline
{\small FM\textsuperscript{(1)}-JD \,(Greedy)} & {\small 11.41(1.06)} & {\small 154.47(17.59)}\tabularnewline
\bottomrule
\end{tabular}}{\small\par}
\end{table}
\begin{table}
\caption{Average error over all MC runs for each node. One standard deviation
shown in parenthesis{\small}.\protect\label{tab:case1-ospa2-per-node}}

\centering{}%
\begin{tabular}{|c|c||c|c|}
\hline 
\noalign{\vskip1bp}
{\small\textbf{Node Loc.}} & {\small\textbf{OSPA\textsuperscript{\textbf{(2)}}(m)}} & {\small\textbf{Node Loc.}} & {\small\textbf{OSPA\textsuperscript{\textbf{(2)}}(m)}}\tabularnewline[1bp]
\hline 
\hline 
{\small (50,950)} & {\small 40.02(7.28)} & {\small (1950,50)} & {\small 41.26(7.35)}\tabularnewline
\hline 
{\small (1000,950)} & {\small 48.54(4.35)} & {\small (1000,50)} & {\small 49.00(5.14)}\tabularnewline
\hline 
{\small (1950,950)} & {\small 41.53(7.30)} & {\small (50,50)} & {\small 39.24(6.87)}\tabularnewline
\hline 
{\small (1950,500)} & {\small 61.56(4.35)} & {\small (50,500)} & {\small 62.09(4.54)}\tabularnewline
\hline 
\end{tabular}
\end{table}

\section{Conclusions\protect\label{sec:Conclusion}}

In this work, we introduced the notion of average for trajectories
and multi-sets of trajectories using the Fr\'{e}chet mean and OSPA-based
metrics. We proposed methods for computing these means based on greedy
search or Gibbs sampling. The greedy-search-based algorithms are numerically
efficient and reach consensus much faster than state-of-the-art algorithms.
In the absence of theoretical results on the convergence to the optimal
solution, we provide empirical analysis by benchmarking the greedy-based
solution against the computationally intensive Gibbs-sampling-based
solutions that have well established convergence properties. Our experiments
on distributed multi-object tracking have shown that the proposed
consensus solutions significantly outperform state-of-the-art techniques.
The greedy search algorithms outperform the current techniques both
in terms of efficiency and accuracy by a significant margin, and are
not far from the Gibbs-sampling-based solution in accuracy. Future
research will focus on developing efficient algorithms for diverse
tracking applications, including scenarios with unknown and non-common
fields of view, where our consensus approach shows significant potential
for adaptation and advancement. 

\bibliographystyle{IEEEtran}
\bibliography{ref_consensus_abbrv_rev,ref_track_fusion_abbrv_rev}

\cleardoublepage{}

\setcounter{page}{1}

\setcounter{section}{0}

\setcounter{equation}{0} 
\renewcommand{\theequation}{A\arabic{equation}}

\noindent\begin{minipage}[c]{1\columnwidth}%
{\large\textbf{Supplementary Materials:}}{\large\par}

{\large\smallskip{}
}{\large\par}

{\large The Mean of Multi-Object Trajectories\medskip{}
}{\large\par}

Tran Thien Dat Nguyen, Ba Tuong Vo, Ba-Ngu Vo,

Hoa Van Nguyen, and Changbeom Shim%
\end{minipage}

\section{Mathematical Proofs\protect\label{sec:Mathematical-Proofs}}

In this section, we provide mathematical proofs that were left out
in the main text.\vspace{-0em}

\subsection{OSPA Trajectory Metric Proof\protect\label{subsec:proof-prop:ospa-track}}

Using the abbreviation $\mathcal{D}_{\!uv}=\mathcal{D}_{\!u}\cup\mathcal{D}_{\!v}$,
we rewrite the OSPA distance (\ref{eq:ospa-track-metric}) between
any two trajectories $u,v\in\mathbb{T}$ as\vspace{-0em}
\[
d_{\mathbb{T}}^{(c,r)}(u,\!v)\!=\!\left[\frac{1}{|\mathcal{D}_{\!uv}|}\sum_{k\in\mathcal{D}_{\!u}\cup\mathcal{D}_{\!v}}[\tilde{d}_{\mathbb{X}}^{(c)}\!(u_{k},v_{k})]^{r}\right]^{\frac{1}{r}},
\]
where $u_{k}=\{u(k)\}$ if $k\in\mathcal{D}_{\!u}$, $u_{k}=\emptyset$
if $k\notin\mathcal{D}_{\!u}$, 
\[
\tilde{d}_{\mathbb{X}}^{(c)}(x,y)=\begin{cases}
d_{\mathbb{X}}^{(c)}(x^{(1)},y^{(1)}), & |x|=|y|=1\\
0, & |x|=|y|=0\\
c, & \textrm{otherwise}
\end{cases},
\]
and $x\!=\!\{x^{(1)}\!,...,x^{(n)}\!\}$. The proof is based on the
following lemma.

\noindent\textbf{Lemma A1.} \label{lem:track-bounds}For any trajectories
$u$, $v$, $t\in\mathbb{T}$, $c>0$ and $r\in[1,\infty]$, ,\vspace{-0.5em}
\[
\left[\frac{1}{|\mathcal{D}_{\!ut}|}\!\sum_{k\in\mathcal{D}_{\!ut}}[\tilde{d}_{\mathbb{X}}^{(c)}(u_{k},v_{k})]^{r}\right]^{\frac{1}{r}}\!\!\!\le\!\!\left[\frac{1}{|\mathcal{D}_{\!uv}|}\!\sum_{k\in\mathcal{D}_{\!uv}}[\tilde{d}_{\mathbb{X}}^{(c)}(u_{k},v_{k})]^{r}\!\right]^{\frac{1}{r}}\!\!.
\]

\noindent\textit{Proof.} For $r=\infty$, the inequality holds because\vspace{-0.5em}
\begin{equation}
\lim_{r\rightarrow\infty}\left[\frac{1}{n}\sum_{i=1}^{n}[\alpha_{i}]^{r}\right]^{\frac{1}{r}}=\max_{1\leq i\leq n}\alpha_{i},\label{eq:lim-r}
\end{equation}
for $\alpha_{i}\geq0$ and $\tilde{d}_{\mathbb{X}}^{(c)}(u_{k},v_{k})=0$
for $k\notin\mathcal{D}_{\!uv}$. 

\noindent For $r\in[1,\infty)$, let $I=\mathcal{D}_{\!uv}\cap\mathcal{D}_{\!ut}$
and $V=\mathcal{D}_{\!uv}\setminus\mathcal{D}_{\!ut}$, which means
$\mathcal{D}_{\!uv}=I\uplus V$. Noting that $|\mathcal{D}_{\!uv}|=|I|+|V|$
and $|I|=|\mathcal{D}_{\!uv}\cap\mathcal{D}_{\!ut}|\leq|\mathcal{D}_{\!ut}|$
we have, 
\begin{align*}
|\mathcal{D}_{\!uv}|\thinspace|I|\leq\thinspace & |\mathcal{D}_{\!ut}|(|I|+|V|)\\
\Leftrightarrow(|\mathcal{D}_{\!uv}|-|\mathcal{D}_{\!ut}|)|I|c^{r}\leq\thinspace & |\mathcal{D}_{\!ut}||V|c^{r}\\
\Leftrightarrow(|\mathcal{D}_{\!uv}|-|\mathcal{D}_{\!ut}|)\sum_{k\in I}[\tilde{d}_{\mathbb{X}}^{(c)}(u_{k},v_{k})]^{r}\leq\thinspace & |\mathcal{D}_{\!ut}||V|c^{r}
\end{align*}
\vspace{-1.5em}
\begin{align*}
\Leftrightarrow\!|\mathcal{D}_{\!uv}|\sum_{k\in I}[\tilde{d}_{\mathbb{X}}^{(c)}(u_{k},v_{k})]^{r}\! & \!\leq\!|\mathcal{D}_{\!ut}|\!\left[\sum_{k\in I}[\tilde{d}_{\mathbb{X}}^{(c)}(u_{k},v_{k})]^{r}\!+|V|c^{r}\right]\\
\!\!\!\!\!\!\Leftrightarrow\!\frac{1}{|\mathcal{D}_{\!ut}|}\!\sum_{k\in I}[\tilde{d}_{\mathbb{X}}^{(c)}(u_{k},v_{k})]^{r}\! & \!\leq\!\frac{1}{|\mathcal{D}_{\!uv}|}\!\!\left[\sum_{k\in I}[\tilde{d}_{\mathbb{X}}^{(c)}(u_{k},v_{k})]^{r}\!\!+|V|c^{r}\!\right]\!\!.
\end{align*}

\noindent Since $\mathcal{D}_{\!ut}=I\uplus(\mathcal{D}_{\!ut}\setminus\mathcal{D}_{\!uv})$
and $\tilde{d}_{\mathbb{X}}^{(c)}(u_{k},v_{k})=0$ for $k\notin\mathcal{D}_{\!uv}$,
\begin{align*}
\sum_{k\in\mathcal{D}_{\!ut}}[\tilde{d}_{\mathbb{X}}^{(c)}(u_{k},v_{k})]^{r}= & \sum_{k\in I}[\tilde{d}_{\mathbb{X}}^{(c)}(u_{k},v_{k})]^{r}.
\end{align*}

\noindent In addition, since $\mathcal{D}_{\!uv}=I\uplus V$ and $[\tilde{d}_{\mathbb{X}}^{(c)}(u_{k},v_{k})]^{r}=c^{r}$
for $k\in V$ (i.e., $k\in\mathcal{D}_{\!v}$, $k\notin\mathcal{D}_{\!u}$,
$k\notin\mathcal{D}_{t}$ ), 
\begin{align*}
\sum_{k\in\mathcal{D}_{\!uv}}[\tilde{d}_{\mathbb{X}}^{(c)}(u_{k},v_{k})]^{r}= & \sum_{k\in I}[\tilde{d}_{\mathbb{X}}^{(c)}(u_{k},v_{k})]^{r}+|V|c^{r}.
\end{align*}
Hence, 
\[
\frac{1}{|\mathcal{D}_{\!ut}|}\sum_{k\in\mathcal{D}_{\!ut}}[\tilde{d}_{\mathbb{X}}^{(c)}(u_{k},v_{k})]^{r}\le\frac{1}{|\mathcal{D}_{\!uv}|}\sum_{k\in\mathcal{D}_{\!uv}}[\tilde{d}_{\mathbb{X}}^{(c)}(u_{k},v_{k})]^{r}.
\]
Raising both sides to $1/r$ yields the desired result. \qed

By definition, (\ref{eq:ospa-track-metric}) trivially satisfies the
metric properties except for the triangle inequality. Thus, for any
trajectories $u$, $v$, $t\in\mathbb{T}$, we need to prove
\begin{equation}
d_{\mathbb{T}}^{(c,r)}(u,v)\leq d_{\mathbb{T}}^{(c,r)}(v,t)+d_{\mathbb{T}}^{(c,r)}(u,t).\label{eq:tri-in-ospa-traj}
\end{equation}

Consider $r\in[1,\infty)$. Since $\tilde{d}_{\mathbb{X}}^{(c)}$
is a metric,
\[
\tilde{d}_{\mathbb{X}}^{(c)}(u_{k},v_{k})\leq\tilde{d}_{\mathbb{X}}^{(c)}(v_{k},t_{k})+\tilde{d}_{\mathbb{X}}^{(c)}(u_{k},t_{k}),
\]
and hence,
\begin{align*}
 & \left[\frac{1}{|\mathcal{D}_{\!uv}|}\sum_{k\in\mathcal{D}_{\!uv}}[\tilde{d}_{\mathbb{X}}^{(c)}(u_{k},v_{k})]^{r}\right]^{\frac{1}{r}}\\
\leq & \left[\frac{1}{|\mathcal{D}_{\!uv}|}\sum_{k\in\mathcal{D}_{\!uv}}[\tilde{d}_{\mathbb{X}}^{(c)}(v_{k},t_{k})+\tilde{d}_{\mathbb{X}}^{(c)}(u_{k},t_{k})]^{r}\right]^{\frac{1}{r}}.
\end{align*}
Using the definition of $d_{\mathbb{T}}^{(c,r)}$ and applying Minkowski's
inequality to the right-hand side, it follows that
\begin{align*}
d_{\mathbb{T}}^{(c,r)}(u,v) & \leq\left[\frac{1}{|\mathcal{D}_{\!uv}|}\sum_{k\in\mathcal{D}_{\!uv}}[\tilde{d}_{\mathbb{X}}^{(c)}(v_{k},t_{k})]^{r}\right]^{\frac{1}{r}}\\
 & +\left[\frac{1}{|\mathcal{D}_{\!uv}|}\sum_{k\in\mathcal{D}_{\!uv}}[\tilde{d}_{\mathbb{X}}^{(c)}(u_{k},t_{k})]^{r}\right]^{\frac{1}{r}}.
\end{align*}
Applying Lemma A1 yields
\begin{align*}
\left[\frac{1}{|\mathcal{D}_{\!uv}|}\sum_{k\in\mathcal{D}_{\!uv}}[\tilde{d}_{\mathbb{X}}^{(c)}(v_{k},t_{k})]^{r}\right]^{\frac{1}{r}}\!\!\! & \leq\!\!\left[\frac{1}{|\mathcal{D}_{\!vt}|}\sum_{k\in\mathcal{D}_{\!vt}}[\tilde{d}_{\mathbb{X}}^{(c)}(v_{k},t_{k})]^{r}\right]^{\frac{1}{r}}\!\!\!,\\
\left[\frac{1}{|\mathcal{D}_{\!uv}|}\sum_{k\in\mathcal{D}_{\!uv}}[\tilde{d}_{\mathbb{X}}^{(c)}(u_{k},t_{k})]^{r}\right]^{\frac{1}{r}}\!\!\! & \leq\!\!\left[\frac{1}{|\mathcal{D}_{\!ut}|}\sum_{k\in\mathcal{D}_{\!ut}}[\tilde{d}_{\mathbb{X}}^{(c)}(u_{k},t_{k})]^{r}\right]^{\frac{1}{r}}\!\!\!.
\end{align*}
Thus, (\ref{eq:tri-in-ospa-traj}) holds by the definition of $d_{\mathbb{T}}^{(c,r)}$
and the above inequality. 

For $r=\infty$, it follows from (\ref{eq:lim-r}) that $\lim_{r\rightarrow\infty}d_{\mathbb{T}}^{(c,r)}(u,v)=d_{\mathbb{T}}^{(c,\infty)}(u,v)$.
Thus, (\ref{eq:tri-in-ospa-traj}) also holds using the same argument
as per the $r\in[1,\infty)$ case.\vspace{-1em}

\subsection{Proof of Proposition \ref{prop:v-opt}\protect\label{subsec:proof-prop:v-opt}}

Suppose (\ref{eq:v_opt}) is not true, i.e., there is an $l\in\mathcal{D}_{\hat{v}}$
and a $\mu\in\mathbb{X}$ such that
\begin{eqnarray*}
{\displaystyle \sum_{n:\mathcal{D}_{\!v^{(n)}\!}\ni l}}\frac{[d_{\mathbb{X}}^{(c)\!}(\mu,v^{(n)\!}(l))]^{r}}{|\mathcal{D}_{\hat{v}}\cup\mathcal{D}_{v^{(n)}}|} & \!\!\!<\!\! & {\displaystyle \sum_{n:\mathcal{D}_{\!v^{(n)}\!}\ni l}}\frac{[d_{\mathbb{X}}^{(c)\!}(\hat{v}(l),v^{(n)\!}(l))]^{r}}{|\mathcal{D}_{\hat{v}}\cup\mathcal{D}_{v^{(n)}}|}.
\end{eqnarray*}
Let $u$ be a trajectory with the same domain as $\hat{v}$, i.e.,
$\mathcal{D}_{\hat{v}}=\mathcal{D}_{u}$, where $u(l)=\mu$, and $u(k)=\hat{v}(k)$
for each $k\in\mathcal{D}_{\hat{v}}-\left\{ l\right\} $. Then, adding
${\displaystyle {\textstyle \sum_{n=1}^{N}}}\mathbf{1}_{\mathcal{\overline{D}}_{\!v^{(n)}\!}}^{(l)}c^{r}|\mathcal{D}_{u}\cup\mathcal{D}_{v^{(n)}}|^{-1}$
to both sides of the above inequality we have
\begin{eqnarray*}
\Psi_{l}^{(r)}(u)\!\! & \!\!\!\!=\!\!\!\! & \!\!{\displaystyle \sum_{n=1}^{N}}\frac{\mathbf{1}_{\mathcal{D}_{\!v^{(n)}\!}}^{(l)}[d_{\mathbb{X}}^{(c)\!}(\mu,v^{(n)\!}(l))]^{r}+\mathbf{1}_{\mathcal{\overline{D}}_{\!v^{(n)}\!}}^{(l)}c^{r}}{|\mathcal{D}_{u}\cup\mathcal{D}_{v^{(n)}}|}\\
\!\! & \!\!\!\!<\!\!\!\! & \!\!{\displaystyle \sum_{n=1}^{N}}\frac{\mathbf{1}_{\mathcal{D}_{\!v^{(n)}\!}}^{(l)}[d_{\mathbb{X}}^{(c)\!}(\hat{v}(l),v^{(n)\!}(l))]^{r}\!+\!\mathbf{1}_{\mathcal{\overline{D}}_{\!v^{(n)}\!}}^{(l)}c^{r}}{|\mathcal{D}_{\hat{v}}\cup\mathcal{D}_{v^{(n)}}|}\!=\!\Psi_{l}^{(r)}(\hat{v}).
\end{eqnarray*}
Further, $\mathbf{1}_{\mathcal{\overline{D}}_{\hat{v}}}^{(k)}\bar{\Psi}_{k}^{(r)}(\hat{v})=\mathbf{1}_{\mathcal{\overline{D}}_{u}}^{(k)}\bar{\Psi}_{k}^{(r)}(u)$
because these terms only depend on $\mathcal{D}_{\hat{v}}=\mathcal{D}_{u}$.
Hence,
\begin{eqnarray*}
\!\!\!\!\!\!V^{(r)}(u) & \!\!\!=\!\!\! & {\displaystyle \sum_{k\in\mathbb{K}}}\mathbf{1}_{\mathcal{D}_{u}}^{(k)}\Psi_{k}^{(r)}(u)+\mathbf{1}_{\mathcal{\overline{D}}_{u}}^{(k)}\bar{\Psi}_{k}^{(r)}(u)\\
 & \!\!\!<\!\!\! & {\displaystyle \sum_{k\in\mathbb{K}}}\mathbf{1}_{\mathcal{D}_{\hat{v}}}^{(k)}\Psi_{k}^{(r)}(\hat{v})+\mathbf{1}_{\mathcal{\overline{D}}_{\hat{v}}}^{(k)}\bar{\Psi}_{k}^{(r)}(\hat{v})=V^{(r)}(\hat{v}).
\end{eqnarray*}
This contradicts the minimality of $\hat{v}$, hence the result.

\subsection{Proof of Lemma \ref{lem:problem-transform-traj}\protect\label{subsec:proof-lem:problem-transform-traj}}

Let $y^{*}\in\mathbb{Y}$ be a minimizer of $g$, and $z^{*}=\mathcal{R}(y^{*})\in\mathbb{Z}$.
Since $\mathcal{R\circ A}=\mathbf{I}$, we have $\mathcal{R}(\mathcal{A}(z^{*}))=z^{*}=\mathcal{R}(y^{*})$,
which implies $g(\mathcal{A}(z^{*}))=g(y^{*})$. Further, using $g\circ\mathcal{A}=f$,
we have $f(z^{*})=g(\mathcal{A}(z^{*}))=g(y^{*})$.

Suppose there exists a $\hat{z}\in\mathbb{Z}$ such that $f(\hat{z})<f(z^{*})$.
Then $g(\mathcal{A}(\hat{z}))=f(\hat{z})<f(z^{*})=g(y^{*})$, which
contradicts the minimality of $y^{*}$. Hence, if $y^{*}$ minimizes
$g$, then $z^{*}=\mathcal{R}(y^{*})$ minimizes $f$, and $f(z^{*})=g(y^{*})$.\vspace{-1em}

\subsection{Proof of Proposition \ref{prop:transformed-problem-traj}\protect\label{subsec:proof-prop:transformed-problem-traj}}

It can be verified that $U^{(r)}\circ\mathcal{A}=V^{(r)}$, by substituting
$(\gamma,x)=\mathcal{A}(u)$, i.e., $\gamma_{k}=\mathbf{1}_{\mathcal{D}_{u}}^{(k)}$,
$x_{k}=\mathbf{1}_{\mathcal{D}_{u}}^{(k)}u(k)$, and $\mathcal{D}_{\gamma}=\mathcal{D}_{u}$
into $U^{(r)}(\gamma,x)$. Additionally, if $\mathcal{R}(\gamma,x)=\mathcal{R}(\gamma',x')$,
then $\gamma=\gamma'$ and $x_{k}=x'_{k}$ whenever $\gamma_{k}=1$,
which implies $\gamma_{k}\psi_{k}^{(r)}(\gamma,x_{k})=\gamma'_{k}\psi_{k}^{(r)}(\gamma',x'_{k})$,
and hence, $U^{(r)}(\gamma,x)=U^{(r)}(\gamma',x')$. Since $\mathcal{R\circ A}=\mathbf{I}$,
by virtue of Lemma \ref{lem:problem-transform-traj}, solving (\ref{eq:trackbary-cost})
amounts to solving (\ref{eq:joint-gamma-x}) and applying the transformation
$\mathcal{R}$ to recover the optimal trajectory.\vspace{-1em}

\subsection{Proof of Proposition \ref{prop:mean-single-state}\protect\label{subsec:proof-prop:mean-single-state}}

Using the definition of $U^{(r)}(\gamma,x)$ in (\ref{eq:Ur}), 
\begin{eqnarray*}
U^{(r)}(\gamma,x^{*}(\gamma)) & \!\!\!\!\!=\!\!\!\!\!\! & \min_{x\in\mathbb{X}^{|\mathbb{K}|}}{\displaystyle \sum_{k\in\mathbb{K}}}\gamma_{k}\psi_{k}^{(r)}(\gamma,x_{k})+\left(1-\gamma_{k}\right)\bar{\psi}_{k}^{(r)}(\gamma)\\
 & \!\!\!\!\!\!=\!\!\!\!\!\! & {\displaystyle \sum_{k\in\mathbb{K}}}\min_{x_{k}\in\mathbb{X}}\gamma_{k}\psi_{k}^{(r)}(\gamma,x_{k})\!+\!{\displaystyle \sum_{k\in\mathbb{K}}}\!\left(1\!-\!\gamma_{k}\right)\!\bar{\psi}_{k}^{(r)}(\gamma).
\end{eqnarray*}
Hence, for each $k$, $x_{k}^{*}(\gamma)=\arg\min_{x_{k}\in\mathbb{X}}\gamma_{k}\psi_{k}^{(r)}(\gamma,x_{k})$.
If $\gamma_{k}\!=\!0$ or $\ \mathbf{1}_{\mathcal{D}_{\!v^{(n)}\!}}^{(k)}=\!0$,
for all $n=\!1:N$, then any value of $x_{k}^{*}(\gamma)\!\in\!\mathbb{X}$
yields the same cost. Thus, $x_{k}^{*}(\gamma)\!=\!\boldsymbol{0}\in\mathbb{X}$
is an optimal solution. If $\gamma_{k}\!=\!1$, then $\gamma_{k}\psi_{k}^{(r)}(\gamma,\cdot)=\psi_{k}^{(r)}(\gamma,\cdot)$,
and \vspace{-1em}
\begin{eqnarray*}
\!\!\!\!\!\!x_{k}^{*}(\gamma) & \!\!\!=\!\!\! & \arg\min_{x_{k}\in\mathbb{X}}\psi_{k}^{(r)}(\gamma,x_{k})\\
 & \!\!\!=\!\!\! & \arg\min_{x_{k}\in\mathbb{X}}{\displaystyle \sum_{n=1}^{N}}\frac{\mathbf{1}_{\mathcal{D}_{\!v^{(n)}\!}}^{(k)}[d_{\mathbb{X}}^{(c)\!}(x_{k},v^{(n)\!}(k))]^{r}}{|\mathcal{D}_{\gamma}\cup\mathcal{D}_{v^{(n)}}|}.
\end{eqnarray*}
 \vspace{-1em}

\subsection{Proof of Proposition \ref{prop:max-ospa-mean-card} \protect\label{subsec:proof-prop-max-ospa-mean-card}}

If $L\triangleq\sum_{n=1}^{N}|\boldsymbol{Y}^{(n)}|=0$, an optimal
solution (of Problem (\ref{eq:mean-multi-trajectory-def})-(\ref{eq:ospa-based-traj-ori}))
is $\hat{\boldsymbol{Y}}=\emptyset$, yielding zero cost. 

We now consider $L>0$. For $p=0$, the cardinality difference has
no contribution to the cost and an optimal solution is $\hat{\boldsymbol{Y}}=\cup_{n=1}^{N}\boldsymbol{Y}^{(n)}$,
yielding zero cost, hence the bound holds. For $p>0$, note that for
each $\boldsymbol{Y}^{(i)}$, the multi-object trajectory $\bar{\boldsymbol{Y}}\triangleq\uplus_{j=1}^{N}\boldsymbol{Y}^{(j)}$
matches every elements of $\boldsymbol{Y}^{(i)}$, and the remaining
unmatched elements from $\boldsymbol{Y}^{(1:i-1)},\boldsymbol{Y}^{(i+1:N)}$
constitute a cost of\vspace{-1em}
\begin{align*}
 & \frac{p^{r}\sum_{j=1,j\neq i}^{N}|\boldsymbol{Y}^{(i)}|}{\kappa(\sum_{i=1}^{N}|\boldsymbol{Y}^{(i)}|)}.
\end{align*}
Hence, the total cost of $\bar{\boldsymbol{Y}}$ is
\begin{align*}
\frac{\sum_{i=1}^{N}\left(p^{r}\sum_{j=1,j\neq i}^{N}|\boldsymbol{Y}^{(j)}|\right)}{\kappa(\sum_{i=1}^{N}|\boldsymbol{Y}^{(i)}|)} & =\frac{(N-1)p^{r}\sum_{i=1}^{N}|\boldsymbol{Y}^{(i)}|}{\kappa(\sum_{i=1}^{N}|\boldsymbol{Y}^{(i)}|)},\\
 & =\frac{(N-1)p^{r}L}{\kappa(L)}.
\end{align*}

Now, suppose $\hat{\boldsymbol{Y}}$ is an optimal solution with cost
$D$, and cardinality $|\hat{\boldsymbol{Y}}|>L$. Then, the optimality
of $\hat{\boldsymbol{Y}}$ implies $D\leq\frac{(N-1)p^{r}L}{\kappa(L)}$,
i.e.,
\begin{align*}
 & D\frac{\kappa(L)}{L}\leq(N-1)p^{r},
\end{align*}
and there exists (at least) a trajectory $u\in\hat{\boldsymbol{Y}}$
that is not paired to any sample trajectory in $\boldsymbol{Y}^{(1:N)}$.
Removing $u$ from $\hat{\boldsymbol{Y}}$ changes the cost by 
\begin{align*}
\delta & =\frac{D\kappa(|\hat{\boldsymbol{Y}}|)-Np^{r}}{\kappa(|\hat{\boldsymbol{Y}}|-1)}-D\\
 & =\frac{D\left[\kappa(|\hat{\boldsymbol{Y}}|)-\kappa(|\hat{\boldsymbol{Y}}|-1)\right]-(N-1)p^{r}-p^{r}}{\kappa(|\hat{\boldsymbol{Y}}|-1)}.
\end{align*}
Since $|\hat{\boldsymbol{Y}}|>L$ and $\kappa(j+1)-\kappa(j)\leq\frac{\kappa(L)}{L}$
for all $j\geq L$, 
\[
D\left[\kappa(|\hat{\boldsymbol{Y}}|)-\kappa(|\hat{\boldsymbol{Y}}|-1)\right]\leq D\frac{\kappa(L)}{L}\leq(N-1)p^{r},
\]
which means $\delta<0$. Thus, there is a multi-object trajectory
of cardinality $|\hat{\boldsymbol{Y}}|-1$ with smaller cost than
$\hat{\boldsymbol{Y}}$, thereby contradicting the optimality of $\hat{\boldsymbol{Y}}$.
Hence, it is not possible to have an optimal solution with cardinality
greater than $L$.

\subsection{Proof of Proposition \ref{prop:set-trajectory-problem-transform}\protect\label{subsec:proof-prop:set-trajectory-problem-transform}}

Denote $\left\lceil a,b\right\rceil \!=\!\max\{a,b\}$, $\left\lfloor a,b\right\rfloor \!=\!\min\{a,b\}$,
substituting (\ref{eq:ospa-based-traj-ori}) into (\ref{eq:mean-multi-trajectory-def})
without assuming the larger cardinality yields
\begin{align*}
\hat{\boldsymbol{Y}} & =\arg\min_{\boldsymbol{X}\in\mathcal{M}(\mathbb{T})}\sum_{n=1}^{N}\min_{\pi^{(n)}\in\prod_{\left\lceil |\boldsymbol{X}_{\!}|,|\boldsymbol{Y}_{\!\!}^{(n)}|\right\rceil }}\!\\
 & \sum_{\ell=1}^{\left\lfloor |\boldsymbol{X}_{\!}|,|\boldsymbol{Y}_{\!\!}^{(n)}|\right\rfloor }\!\frac{[\bar{d}_{\mathbb{T}}^{(c)}\!(\boldsymbol{X}_{\!},\boldsymbol{Y}_{\!\!}^{(n)}|\ell,\pi^{(n)})]^{r}\!}{\kappa(|\boldsymbol{X}|,|\boldsymbol{Y}_{\!\!}^{(n)}|)}+\frac{||\boldsymbol{Y}^{(n)\!}|\!-\!|\boldsymbol{X}||p^{r}}{\kappa(|\boldsymbol{X}|,|\boldsymbol{Y}_{\!\!}^{(n)}|)}\!,
\end{align*}
where 
\[
\bar{d}_{\mathbb{T}}^{(c)}\!(\boldsymbol{X}_{\!},\boldsymbol{Y}_{\!\!}|\ell,\pi)=\begin{cases}
d_{\mathbb{T}}^{(c)}\!(\boldsymbol{X}_{\!\ell},\boldsymbol{Y}_{\!\!\pi(\ell)}), & |\boldsymbol{X}|\leq|\boldsymbol{Y}|\\
d_{\mathbb{T}}^{(c)}\!(\boldsymbol{X}_{\!\pi(\ell)},\boldsymbol{Y}_{\ell}), & |\boldsymbol{X}|>|\boldsymbol{Y}|
\end{cases}.
\]
For some $L\geq\left\lceil |\boldsymbol{X}|,|\boldsymbol{Y}^{(n)\!}|\right\rceil $,
we define $\varepsilon:\mathcal{M}(\mathbb{T})\rightarrow\mathbb{B}^{L}$
such that for a multi-object trajectory $\boldsymbol{X}$, $\varepsilon_{\ell}(\boldsymbol{X})=1$
for $\ell\leq|\boldsymbol{X}|$ and $\varepsilon_{\ell}(\boldsymbol{X})=0$
for $\ell>|\boldsymbol{X}|$, $\varepsilon_{\ell}(\boldsymbol{X})$
is the $\ell^{th}$ component of $\varepsilon(\boldsymbol{X})$ and
$\varepsilon(\boldsymbol{X})$ is the existence list of $\boldsymbol{X}$.
We also define a space of permutation $\tilde{\Pi}_{L}^{(n)}(\varepsilon(\boldsymbol{X}))\subseteq\Pi_{L}$
such that for each $\tilde{\pi}\in\tilde{\Pi}_{L}^{(n)}(\varepsilon(\boldsymbol{X}))$
\begin{itemize}
\item $\varepsilon_{\ell}(\boldsymbol{X})=1\Rightarrow\varepsilon_{\tilde{\pi}(\ell)}(\boldsymbol{Y}^{(n)\!})=1$
, if $\left\Vert \varepsilon(\boldsymbol{X})\right\Vert _{1}\leq||\varepsilon(\boldsymbol{Y}^{(n)\!})||_{1}$
; or 
\item $\varepsilon_{\tilde{\pi}(\ell)}(\boldsymbol{Y}^{(n)\!})=1\Rightarrow\varepsilon_{\ell}(\boldsymbol{X})=1$
, if $\left\Vert \varepsilon(\boldsymbol{X})\right\Vert _{1}>||\varepsilon(\boldsymbol{Y}^{(n)\!})||_{1}$
.
\end{itemize}
We obtain the equivalence
\begin{align*}
 & \min_{\pi\in\prod_{\left\lceil |\boldsymbol{X}_{\!}|,|\boldsymbol{Y}_{\!\!}^{(n)}|\right\rceil }}\!\sum_{\ell=1}^{\left\lfloor |\boldsymbol{X}_{\!}|,|\boldsymbol{Y}_{\!\!}^{(n)}|\right\rfloor }\![\bar{d}_{\mathbb{T}}^{(c)}\!(\boldsymbol{X}_{\!},\boldsymbol{Y}^{(n)}|\ell,\pi)]^{r}\\
= & \min_{\tilde{\pi}\in\tilde{\Pi}_{L}^{(n)}(\varepsilon(\boldsymbol{X}))}\sum_{\ell=1}^{L}\thinspace[\tilde{d}_{\mathbb{T}}^{(c)}\!(\boldsymbol{X}_{\!},\boldsymbol{Y}^{(n)}|\ell,\tilde{\pi})]^{r},\textrm{ and}
\end{align*}
\[
||\boldsymbol{Y}^{(n)\!}|-|\boldsymbol{X}||\!\!=\!\!\sum_{\ell=1}^{L}\varepsilon_{\ell}(\boldsymbol{X})(1-\varepsilon_{\ell}(\boldsymbol{Y}^{(n)\!}))+(1-\varepsilon_{\ell}(\boldsymbol{X}))\varepsilon_{\ell}(\boldsymbol{Y}^{(n)\!}),
\]
where
\[
\tilde{d}_{\mathbb{T}}^{(c)}\!(\boldsymbol{X},\boldsymbol{Y}|\ell,\tilde{\pi})=\begin{cases}
d_{\mathbb{T}}^{(c)}\!(\boldsymbol{X}_{\!\ell},\boldsymbol{Y}_{\!\!\tilde{\pi}(\ell)}), & \text{\ensuremath{\ell}\ensuremath{\leq|\boldsymbol{X}|},\ensuremath{\tilde{\pi}}(\ensuremath{\ell})\ensuremath{\leq}|\ensuremath{\boldsymbol{Y}}|}\\
0, & \textrm{otherwise}
\end{cases}.
\]
Further, if we fix $L=\sum_{n=1}^{N}|\boldsymbol{Y}^{(n)}|$ then
$\tilde{\Pi}_{L}^{(n)}(\varepsilon(\boldsymbol{X}))=\Omega^{(n)}(\varepsilon(\boldsymbol{X}))$.
Recall $(\xi^{(n)},\chi^{(n)})=\mathcal{A}(\boldsymbol{Y}^{(n)\!})$
and note that $\varepsilon(\boldsymbol{Y}^{(n)\!})=\xi^{(n)}$, the
original optimization problem becomes
\[
\hat{\boldsymbol{Y}}=\arg\min_{\boldsymbol{X}\in\mathcal{M}(\mathbb{T})}\min_{\omega\in\Omega^{(1:N)}(\varepsilon(\boldsymbol{X}))}P_{\omega}^{(r)}(\boldsymbol{X}),
\]
where
\begin{align*}
P_{\omega}^{(r)}(\boldsymbol{X})= & \sum_{n=1}^{N}\sum_{\ell=1}^{L}\!\frac{[\tilde{d}_{\mathbb{T}}^{(c)}\!(\boldsymbol{X},\boldsymbol{Y}_{\!\!}^{(n)}|\ell,\omega(n,\cdot))]^{r}}{\kappa(|\boldsymbol{X}|,|\boldsymbol{Y}_{\!\!}^{(n)}|)}\\
 & +\frac{\varepsilon_{\ell}(\boldsymbol{X})\!(1\!-\!\xi_{\omega(n,\ell)}^{(n)})p^{r}}{\kappa(|\boldsymbol{X}|,|\boldsymbol{Y}_{\!\!}^{(n)}|)}+\frac{(1-\varepsilon_{\ell}(\boldsymbol{X}))\xi_{\omega(n,\ell)}^{(n)}p^{r}}{\kappa(|\boldsymbol{X}|,|\boldsymbol{Y}_{\!\!}^{(n)}|)}.
\end{align*}
Let $Q^{(r)}(\eta,\tau)=\min_{\omega\in\Omega^{(1:N)}(\eta)}Q_{\omega}^{(r)}(\eta,\tau)$
and $P^{(r)}(\boldsymbol{X})=\min_{\omega\in\Omega^{(1:N)}(\varepsilon(\boldsymbol{X}))}P_{\omega}^{(r)}(\boldsymbol{X})$.
It can be verified that $Q^{(r)}\circ\mathcal{A}=P^{(r)}$, by substituting
$(\eta,\tau)=\mathcal{A}(\boldsymbol{X})$ into the expression of
$P^{(r)}$, i.e., $\varepsilon(\boldsymbol{X})=\eta$, $|\boldsymbol{X}|=\left\Vert \eta\right\Vert _{1}$,
$\boldsymbol{X}_{\ell}=\tau_{\ell}$, and $\tilde{d}_{\mathbb{T}}^{(c)}\!(\boldsymbol{X},\boldsymbol{Y}_{\!\!}^{(n)}|\ell,\omega(n,\cdot))=d_{\mathbb{T}}^{(c)}\!(\boldsymbol{X}_{\!\ell},\boldsymbol{Y}_{\!\!\omega(n,\ell)}^{(n)})=d_{\mathbb{T}}^{(c)}\!(\tau_{\ell},\chi_{\omega(n,\ell)}^{(n)})>0$
only when $\varepsilon_{\ell}(\boldsymbol{X})=1$ and $\varepsilon_{\omega(n,\ell)}(\boldsymbol{Y}^{(n)\!})=1$.
Additionally, if $\mathcal{R}(\eta,\tau)=\mathcal{R}(\eta',\tau')$,
then $\left\Vert \eta\right\Vert _{1}=\left\Vert \eta'\right\Vert _{1}$
and $\{\tau_{\ell}|\eta_{\ell}=1,\ell=1:L\}=\{\tau'_{\ell}|\eta'_{\ell}=1,\ell=1:L\}$,
which implies $Q^{(r)}(\eta,\tau)=Q^{(r)}(\eta',\tau')$. Since $\mathcal{R\circ A}=\mathbf{I}$,
by the virtue of Lemma \ref{lem:problem-transform-traj}, solving
(\ref{eq:mean-multi-trajectory-def}) with the distance form given
by (\ref{eq:ospa-based-traj-ori}) amounts to solving (\ref{eq:multi-traj-mean-problem-trans})
and applying the transformation $\mathcal{R}$ to recover the optimal
multi-object trajectory.

\subsection{Proof of Proposition \ref{prop:mean-single-trajectory}\protect\label{subsec:proof-prop:set-trajectory-problem-transform-2}}

Using the definition of $Q_{\omega}^{(r)}(\eta,\tau)$ in (\ref{eq:cost-multi-traj-mean-problem-trans}),
\begin{align*}
Q_{\omega}^{(r)}(\eta,\tau^{*}(\eta,\omega)) & \!=\!\min_{\tau\in\mathbb{T}^{L}}\sum_{\ell=1}^{L}\eta_{\ell}\phi_{\omega,\ell}^{(r)}(\eta,\tau_{\ell})\text{+}(1-\eta_{\ell})\bar{\phi}_{\omega,\ell}^{(r)}(\eta)\\
 & \!=\!\sum_{\ell=1}^{L}\min_{\tau_{\ell}\in\mathbb{T}}\eta_{\ell}\phi_{\omega,\ell}^{(r)}(\eta,\tau_{\ell})\text{+}\!\sum_{\ell=1}^{L}(1\!-\!\eta_{\ell})\bar{\phi}_{\omega,\ell}^{(r)}(\eta).
\end{align*}
For each $\ell$, $\tau_{\ell}^{*}(\eta,\omega)=\arg\min_{\tau_{\ell}\in\mathbb{T}}\eta_{\ell}\phi_{\omega,\ell}^{(r)}(\eta,\tau_{\ell})$.
If $\eta_{\ell}=0$ or $\xi_{\omega(n,\ell)}^{(n)}=0$ for all $n=\!1\!:\!N$,
any value of $\tau_{\ell}^{*}(\eta,\omega)$ yields the same cost.
Hence, $\tau_{\ell}^{*}(\eta,\omega)\!=\!\boldsymbol{0}$ is an optimal
solution. If $\eta_{\ell}=1$, then $\eta_{\ell}\phi_{\omega,\ell}^{(r)}(\eta,\tau_{\ell})=\phi_{\omega,\ell}^{(r)}(\eta,\tau_{\ell})$,
and
\begin{align*}
\tau_{\ell}^{*}(\eta,\omega) & =\arg\min_{\tau_{\ell}\in\mathbb{T}}\phi_{\omega,\ell}^{(r)}(\eta,\tau_{\ell})\\
 & =\arg\min_{\tau_{\ell}\in\mathbb{T}}\sum_{n=1}^{N}\xi_{\omega(n,\ell)}^{(n)}\frac{d_{\mathbb{T}}^{(c)}(\tau_{\ell},\chi_{\omega(n,\ell)}^{(n)})^{r}}{\kappa(|\!|\eta|\!|_{1},|\!|\xi^{(n)}|\!|_{1})}.
\end{align*}

\section{Extensions on Consensus via Gibbs Sampling}

This section provides a mathematical proof for the convergence of
the Gibbs sampler proposed in Subsection \ref{sec:Gibbs-mo-trajectory},
and discusses efficient implementation strategies.

\subsection{Convergence Properties\protect\label{subsec:Convergence-Property}}

It is not obvious that the Gibbs sampler in Algorithm \ref{alg:bary-ospa-samp}
can visit all possible states in a finite number of steps. However,
this is demonstrated in the following proposition.

\noindent\textbf{Proposition A1.} \label{prop:gibbs-set-trajectory}Starting
from any valid $(\eta_{1:L},\omega^{(1:N)})$ the Gibbs sampler defined
by the conditionals $\rho_{\ell}(\cdot|\acute{\eta}_{1:\ell-1},\eta_{\ell+1:L},\omega^{(1:N)})$
and $\rho_{L+n}(\cdot|\acute{\eta}_{1:L},\acute{\omega}^{(1:n-1)},\omega^{(n+1:N)})$
converges to the distribution $\rho(\eta_{1:L},\omega^{(1:N)})$ at
an exponential rate.

\begin{proof_dat}We need to show that the chain can move from any
valid state $(\eta_{1:L},\omega^{(1:N)})$ to any other valid state
$(\acute{\eta}_{1:L},\acute{\omega}^{(1:N)})$ in a finite number
of steps. Noting that $\rho(\eta_{1:L},\omega^{(1:N)})$ is zero for
invalid samples, i.e., when $\omega^{(n)}\notin\Omega^{(n)}(\eta_{1:L})$
for any $n\in\{1:N\}$, and is positive otherwise. Thus, we can transit
from any valid $(\eta_{1:L},\omega^{(1:N)})$ to any other valid $(\eta_{1:L},\acute{\omega}^{(1:N)})$
in a finite number of steps. We also demonstrate that it is possible
to move from $\eta_{1:L}$ to $\acute{\eta}_{1:L}$ in a finite number
of steps. For any $\ell\in\{1:L\}$ and $n\in\{1:N\}$, we proceed
with the following cases:

\textit{Case 1:} Suppose $\left\Vert \eta_{1:L}\right\Vert _{1}=\left\Vert \xi^{(n)}\right\Vert _{1}$.
Changing $\eta_{\ell}$ from 0 to 1 (or vice versa) to reach some
intermediate state $\grave{\eta}_{1:L}$ will always satisfy $\omega^{(n)}\in\Omega^{(n)}(\grave{\eta}_{1:L})$.
This permits a transition from $(\eta_{1:L},\omega^{(1:N)})$ to the
intermediate state $(\acute{\eta}_{1:L},\omega^{(1:N)})$ and then
$(\acute{\eta}_{1:L},\acute{\omega}^{(1:N)})$.

\textit{Case 2:} Suppose $\left\Vert \eta_{1:L}\right\Vert _{1}<\left\Vert \xi^{(n)}\right\Vert _{1}$.
Changing $\eta_{\ell}$ from 1 to 0 to reach an intermediate state
$\grave{\eta}_{1:L}$ will satisfy $\omega^{(n)}\in\Omega^{(n)}(\grave{\eta}_{1:L})$,
allowing a transition from $(\eta_{1:L},\omega^{(1:N)})$ to $(\grave{\eta}_{1:L},\omega^{(1:N)})$.
If $\xi_{\ell}^{(n)}=1$, changing $\eta_{\ell}$ from 0 to 1 to reach
$\grave{\eta}_{1:L}$ will also satisfy $\omega^{(n)}\in\Omega^{(n)}(\grave{\eta}_{1:L})$.
However, if $\xi_{\ell}^{(n)}=0$, to change $\eta_{\ell}$ from 0
to 1 we first need to change $\omega^{(n)}$ to $\grave{\omega}^{(n)}$
as follows. For some $\grave{\ell}$ where $\eta_{\grave{\ell}}=0$
and $\xi_{\grave{\ell}}^{(n)}=1$, we select some $\grave{\omega}^{(n)}\in\Omega^{(n)}(\eta_{1:L})$
such that: $\grave{\omega}_{\ell}^{(n)}=\omega_{\grave{\ell}}^{(n)}$;
$\grave{\omega}_{\grave{\ell}}^{(n)}=\omega_{\ell}^{(n)}$; and $\grave{\omega}_{\check{\ell}}^{(n)}=\omega_{\check{\ell}}^{(n)}$
for $\check{\ell}\neq\ell,\grave{\ell}$. Since $\left\Vert \eta_{1:L}\right\Vert _{1}<\left\Vert \xi^{(n)}\right\Vert _{1}$,
there always exists a $\grave{\ell}$ that satisfies our requirement.
Since $\grave{\omega}^{(n)}\in\Omega^{(n)}(\grave{\eta}_{1:L})$,
it allows us to move from $(\eta_{1:L},\omega^{(1:N)})$ to $(\eta_{1:L},\grave{\omega}^{(1:N)})$
to $(\acute{\eta}_{1:L},\grave{\omega}^{(1:N)})$ and then $(\acute{\eta}_{1:L},\acute{\omega}^{(1:N)})$.

\textit{Case 3}: Suppose $\left\Vert \eta_{1:L}\right\Vert _{1}>\left\Vert \xi^{(n)}\right\Vert _{1}$.
Changing $\eta_{\ell}$ from 0 to 1 to reach an intermediate state
$\grave{\eta}_{1:L}$ will satisfy $\omega^{(n)}\in\Omega^{(n)}(\grave{\eta}_{1:L})$,
allowing a transition from $(\eta_{1:L},\omega^{(1:N)})$ to $(\grave{\eta}_{1:L},\omega^{(1:N)})$.
If $\xi_{\ell}^{(n)}=0$, changing $\eta_{\ell}$ from 1 to 0 to reach
$\grave{\eta}_{1:L}$ will also satisfy $\omega^{(n)}\in\Omega^{(n)}(\grave{\eta}_{1:L})$.
However, if $\xi_{\ell}^{(n)}=1$, to change $\eta_{\ell}$ from 1
to 0 we first need to change $\omega^{(n)}$ to $\grave{\omega}^{(n)}$
as follows. For some $\grave{\ell}$ where $\eta_{\grave{\ell}}=1$
and $\xi_{\grave{\ell}}^{(n)}=0$, we select some $\grave{\omega}^{(n)}\in\Omega^{(n)}(\eta_{1:L})$
such that: $\grave{\omega}_{\ell}^{(n)}=\omega_{\grave{\ell}}^{(n)}$;
$\grave{\omega}_{\grave{\ell}}^{(n)}=\omega_{\ell}^{(n)}$; and $\grave{\omega}_{\check{\ell}}^{(n)}=\omega_{\check{\ell}}^{(n)}$
for $\check{\ell}\neq\ell,\grave{\ell}$. Since $\left\Vert \eta_{1:L}\right\Vert _{1}>\left\Vert \xi^{(n)}\right\Vert _{1}$,
there always exists a $\grave{\ell}$ that satisfies our requirement.
Since $\grave{\omega}^{(n)}\in\Omega^{(n)}(\grave{\eta}_{1:L})$,
it allows us to move from $(\eta_{1:L},\omega^{(1:N)})$ to $(\eta_{1:L},\grave{\omega}^{(1:N)})$
to $(\acute{\eta}_{1:L},\grave{\omega}^{(1:N)})$ and then $(\acute{\eta}_{1:L},\acute{\omega}^{(1:N)})$. 

This implies the Markov chain is irreducible and regular, hence the
convergence to the target distribution $\rho(\eta_{1:L},\omega^{(1:N)})$
at an exponential rate, following from Lemma 4.3.4 of \cite{gallager2013stochastic}.\end{proof_dat}

\subsection{Enhancement Strategies\protect\label{subsec:Enhancement-Strategies}}

In this section, we introduce two techniques to respectively improve
the convergence rate and the efficiency of the Gibbs sampler proposed
in Subsection \ref{sec:Gibbs-mo-trajectory}.

\subsubsection{Rearranging Non-Existence Assignment}

For an existence array $\eta_{1:L}$ and some existence assignments
$\omega^{(1:N)},\grave{\omega}^{(1:N)}\in\Omega^{(1:N)}(\eta_{1:L})$,
if $\omega_{\ell}^{(n)}=\grave{\omega}_{\ell}^{(n)}$ for all $\ell$
that $\eta_{\ell}=1$ then $S^{(r)}(\eta_{1:L},\omega^{(1:N)})=S^{(r)}(\eta_{1:L},\grave{\omega}^{(1:N)})$,
i.e., $\omega^{(1:N)}$ and $\grave{\omega}^{(1:N)}$ yield the same
cost value if they only differ at the components where $\eta_{\ell}=0$
(assignment of the non-existence elements). If $\eta_{\ell}=0$ and
the pair-wise distances among $\chi_{\grave{\omega}_{\ell}^{(1)}}^{(n)}$,
..., $\chi_{\grave{\omega}_{\ell}^{(N)}}^{(n)}$ are lower than ones
among $\chi_{\omega_{\ell}^{(1)}}^{(n)}$, ..., $\chi_{\omega_{\ell}^{(N)}}^{(n)}$,
replacing $\omega^{(1:N)}$ with $\grave{\omega}^{(1:N)}$ could increase
the probability of $\eta_{\ell}$ transiting from 0 to 1 because their
distances to the mean would be shorter.

For some $\eta_{1:L}$, $\bar{L}=\sum_{\ell=1}^{L}(1-\eta_{\ell})$,
we define the mapping $\bar{\mathcal{E}}:\Pi_{L}\rightarrow\mathcal{F}_{\bar{L}}(\{1:L\})$
(where $\mathcal{F}_{\bar{L}}(\{1:L\})$ is all finite subsets of
$\{1:L\}$ with $\bar{L}$ elements) such that $\bar{\mathcal{E}}(\omega)=\{\omega_{\ell}|\eta_{\ell}=0,\ell=1:L\}$.
For some $\omega\in\Pi_{L}$, we define the mapping $\mathcal{E}_{\omega}:\mathcal{F}_{\bar{L}}(\{1:L\})\rightarrow\Pi_{L}$
such that if $\mathcal{E}_{\omega}(\{\bar{\omega}_{1:\bar{L}}\})=\grave{\omega}$
then
\[
\begin{cases}
\grave{\omega}_{\ell}=\omega_{\ell}, & \textrm{if }\eta_{\ell}=1\\
\grave{\omega}_{\ell}=\bar{\omega}_{m},m=\sum_{i=1:\ell}(1-\eta_{i}), & \textrm{if }\eta_{\ell}=0
\end{cases}.
\]
We note that if $\bar{\omega}\in\Pi_{\bar{\mathcal{E}}(\omega)}$
is a permutation of $\bar{\mathcal{E}}(\omega)$ and $\omega\in\Omega^{(n)}(\eta_{1:L})$
then $\mathcal{E}_{\omega}(\bar{\omega})\in\Omega^{(n)}(\eta_{1:L})$.
Thus, we can update $\omega^{(n)}\in\Omega^{(n)}(\eta_{1:L})$ to
$\grave{\omega}^{(n)}\in\Omega^{(n)}(\eta_{1:L})$ using some $\bar{\omega}^{(n)}\in\Pi_{\bar{\mathcal{E}}(\omega^{(n)})}$
such that $\acute{\omega}_{\ell}^{(n)}=\mathcal{E}_{\omega^{(n)}}(\bar{\omega}^{(n)})$.
Noting that $S^{(r)}(\eta_{1:L},\omega^{(1:N)})=S^{(r)}(\eta_{1:L},\grave{\omega}^{(1:N)})$.
To achieve low pair-wise distances among $\chi_{\grave{\omega}_{\ell}^{(1)}}^{(n)}$,
..., $\chi_{\grave{\omega}_{\ell}^{(N)}}^{(n)}$, we sample $\bar{\omega}^{(1:N)}$
from the distribution
\[
\bar{\rho}(\bar{\omega}^{(1:N)})\propto e^{-\alpha(R^{(r)}(\bar{\omega}^{(1:N)}))}\prod_{n=1}^{N}\boldsymbol{1}_{\Pi_{\bar{\mathcal{E}}(\omega^{(n)})}}(\bar{\omega}^{(n)}),
\]
where
\begin{align*}
R^{(r)}(\bar{\omega}^{(1:N)})\!=\! & \sum_{\ell=1}^{\bar{L}}\!\zeta_{\ell}^{(r)}(\bar{\omega}^{(1:N)}),\\
\zeta_{\ell}^{(r)}(\bar{\omega}^{(1:N)})\!=\! & \sum_{n=1}^{N}\sum_{j=1,j\neq n}^{N}\!\xi_{\bar{\omega}_{\ell}^{(n)}}^{(r)}\xi_{\bar{\omega}_{\ell}^{(j)}}^{(j)}d_{\mathbb{T}}^{(c)}\!(\chi_{\bar{\omega}_{\ell}^{(n)}}^{(n)},\chi_{\bar{\omega}_{\ell}^{(j)}}^{(j)})^{r}\\
 & +\xi_{\bar{\omega}_{\ell}^{(n)}}^{(n)}(1-\xi_{\bar{\omega}_{\ell}^{(j)}}^{(j)})p^{r}+(1-\xi_{\bar{\omega}_{\ell}^{(n)}}^{(n)})\xi_{\bar{\omega}_{\ell}^{(j)}}^{(j)}p^{r},
\end{align*}
Sampling from $\bar{\rho}$ can be done using the Gibbs sampling technique
proposed in Subsection \ref{sec:Gibbs-mo-trajectory} of the main
text.

\subsubsection{Efficient Proposal Distribution}

We observe that if the trajectories in $\boldsymbol{Y}^{(1:N)}$ are
assigned to the same trajectory $\boldsymbol{X}_{\ell}$ in the mean
multi-object trajectory, they are expected to be close to each other.
Otherwise, if $\boldsymbol{Y}^{(n)}$ has no trajectories assigned
to $\boldsymbol{X}_{\ell}$, it contributes $p^{r}$ to the cost function.
Therefore, we propose the following proposal conditional $\tilde{\rho}_{L+n}$
in place of $\rho_{L+n}$ in Subsection \ref{sec:Gibbs-mo-trajectory}
of the main text to sample for the existence assignment, i.e.,
\begin{multline*}
\tilde{\rho}_{L+n}(\omega^{(n)}|\acute{\eta}_{1:L},\acute{\omega}^{(1:n-1)},\omega^{(n+1:N)})=\\
e^{-\alpha(\tilde{S}^{(r,n)}(\acute{\eta}_{1:L},[\acute{\omega}^{(1:n-1)},\omega^{(n:N)}]))}\boldsymbol{1}_{\Omega^{(n)}(\eta_{1:L})}(\omega^{(n)}),
\end{multline*}
where
\begin{align*}
\tilde{S}^{(r,n)}(\eta,\omega)= & \sum_{\ell=1}^{L}\eta_{\ell}\acute{\phi}_{\omega,\ell}^{(r,n)}(\eta)\text{+}(1\!-\!\eta_{\ell})\grave{\phi}_{\omega,\ell}^{(r,n)},\\
\acute{\phi}_{\omega,\ell}^{(r,n)}(\eta)= & \!\sum_{j=1,j\neq n}^{N}\!\xi_{\omega_{\ell}^{(n)}}^{(n)}\xi_{\omega_{\ell}^{(j)}}^{(j)}[d_{\mathbb{T}}^{(c)}(\chi_{\omega_{\ell}^{(n)}}^{(n)},\chi_{\omega_{\ell}^{(j)}}^{(j)})]^{r}\\
 & +\xi_{\omega_{\ell}^{(n)}}^{(n)}(1-\xi_{\omega_{\ell}^{(j)}}^{(j)})p^{r}+(1-\xi_{\omega_{\ell}^{(n)}}^{(n)})\xi_{\omega_{\ell}^{(j)}}^{(j)}p^{r},\\
\grave{\phi}_{\omega,\ell}^{(r,n)}= & \sum_{j=1,j\neq n}^{N}\xi_{\omega_{\ell}^{(j)}}^{(j)}p^{r}.
\end{align*}
Compared to $\rho_{L+n}$, we do not need to compute the mean trajectory
$\tau_{\ell}^{*}$ when computing $\tilde{\rho}_{L+n}$. Thus, the
complexity of Algorithm \ref{alg:bary-ospa-samp} reduces to $\mathcal{O}(G_{m}L^{2}Z)$.
\end{document}